\documentclass[fleqn,usenatbib]{mnras}

\usepackage[T1]{fontenc}

\DeclareRobustCommand{\VAN}[3]{#2}
\let\VANthebibliography\thebibliography
\def\thebibliography{\DeclareRobustCommand{\VAN}[3]{##3}\VANthebibliography}

\usepackage{graphicx}	
\usepackage{amsmath}	
\usepackage{amssymb}	
\usepackage{bm}
\usepackage{mathtools}
\usepackage[normalem]{ulem} 

\newcommand{\rhog}{{\rho_{\rm g}}}

\newcommand{\vg}{{\bf{v}_{\rm g}}}

\newcommand{\hatvgx}{{\hat{v}_{{\rm g}x}}}

\newcommand{\hatsigma}{{\hat{\sigma}}}

\newcommand{\ud}{{\bf{u}}}

\newcommand{\taus}{{\tau_{\rm s}}}

\newcommand{\tauspeak}{{\tau_{\rm s,peak}}}
\newcommand{\tausmin}{{\tau_{\rm s,min}}}

\defcitealias{paper1}{Paper~I}
\defcitealias{paper2}{Paper~II}
\defcitealias{2005ApJ...620..459Y}{YG05}
\defcitealias{2019ApJ...878L..30K}{K+19}

\newcounter{foo}

\title[Polydisperse Streaming Instability III]{Polydisperse Streaming Instability III. Dust evolution encourages fast instability}

\author[C.~P. McNally et al.]{
Colin P.~McNally,$^{1}$
\thanks{E-mail: colin@colinmcnally.ca, 
  f.lovascio@qmul.ac.uk,
  s.j.paardekooper@qmul.ac.uk}
\setcounter{foo}{\value{footnote}}
\thanks{Present address: 81 Concession 8~E, Freelton, ON, L8B~1N9, Canada}
Francesco Lovascio,$^{1}$\footnotemark[\value{foo}]
Sijme-Jan Paardekooper$^{1,2}$\footnotemark[\value{foo}]
\\
$^{1}$Astronomy Unit, School of Physics and Astronomy, Queen Mary University of London, London E1 4NS, UK\\
$^{2}$DAMTP, University of Cambridge, Wilberforce Road, Cambridge CB3 0WA, UK
}

\date{Accepted XXX. Received YYY; in original form ZZZ}

\pubyear{2020}

\begin{document}
\label{firstpage}
\pagerange{\pageref{firstpage}--\pageref{lastpage}}
\maketitle

\begin{abstract}
Planet formation via core accretion requires the production of km-sized planetesimals from cosmic dust. This process must overcome barriers to simple collisional growth, for which the Streaming Instability (SI) is often invoked.  Dust evolution is still required to create particles large enough to undergo vigorous instability. The SI has been studied primarily with single size dust, and the role of the full evolved dust distribution is largely unexplored. We survey the Polydispserse Streaming Instability (PSI) with physical parameters corresponding to plausible conditions in protoplanetary discs. We consider a full range of particle stopping times, generalized dust size distributions, and the effect of turbulence. We find that, while the PSI grows in many cases more slowly with a interstellar power-law dust distribution than with a single size, reasonable collisional dust evolution, producing an enhancement of the largest dust sizes, produces instability behaviour similar to the monodisperse case. Considering turbulent diffusion the trend is similar. We conclude that if fast linear growth of PSI is required for planet formation, then dust evolution producing a distribution with peak stopping times on the order of 0.1 orbits and an enhancement of the largest dust significantly above the single power-law distribution produced by a fragmentation cascade is sufficient, along with local enhancement of the dust to gas volume mass density ratio to order unity.
\end{abstract}




\section{Introduction}

In the process of planet formation in a protoplanetary disc, dust grows from individual grains into larger agglomerations.
If planets are to be formed by core accretion, that is through the forming of a self-gravitating solid body as opposed to a gas-dominated self-gravitating collapse, then a chain of mechanisms for growth from the $\mu$m scale to the self-gravitating km scale must be established.
Problematically, between the mm scale and km scales a number of challenges to the growth of solids are present.
As dust agglomerations grow, the speed of their collisions does too.
Given the finite and material dependent bonding forces between the dust monomers, these collisions are eventually become too energetic as the particles grow, and result in breaking the agglomerates back down to smaller fragments, preventing further growth.

If the drag timescale of dust against the gas grows towards the orbital timescale, a second barrier is encountered, because 
the timescale for the gas drag on the dust to decay its orbit and drag it into the star reduces.
This introduces a problem, in that even if the dust collisions at this scale are survivable for the dust in question, the time available to have these collisions and grow into large dust is constrained.
Thus, once the dust cannot grow faster than it drifts, it cannot grow to a large enough size for the radial drift to slow down, allowing it to survive.
Furthermore, the scale of the dust collisions velocities is the same order as the inwards dust drift velocity, so this regime is also the least favourable for surviving fragmentation. For a review of these processes, see e.g. \cite{2014prpl.conf..339T}.

To solve these problems the Streaming Instability \citep[SI,][]{2005ApJ...620..459Y} has been a popular proposal.
If dust can be grown both large enough to have long stopping times, and concentrated abundantly enough, the SI can create dust concentrations able to collapse directly into km-scale self-gravitating planetesimals.
On the path to planet formation, these planetesimals can survive dust collisions, and are large enough to be relatively safe from aerodynamic drag decaying their orbits.

Other processes have been proposed for assembling planetesimals from dust.
For example, turbulent clustering of dust, leading directly to gravitational collapse \citep{2001ApJ...546..496C,2008ApJ...687.1432C,2010Icar..208..518C,2010Icar..208..505C,2016MNRAS.456.2383H,2020ApJ...892..120H}, or the concentration of self-gravitating clumps of dust in pressure traps or zonal flows \citep{
    1972fpp..conf..211W,2009ApJ...697.1269J,2014ApJ...796...31B,
    2018A&A...617A.117R},
vortices \citep{1995A&A...295L...1B,
    2008A&A...491L..41L},
    and secular gravitational instability \citep{    1998Icar..133..298S,2011ApJ...731...99Y,2012ApJ...746...35M,2014ApJ...794...55T,2018PASJ...70....3T,2020ApJ...900..182T} 
    are alternate routes to planetesimal formation.
The difference in the underlying physics between these planetesimal formation scenarios leads to different requirements placed on the dust evolution processes in the molecular cloud, star formation, and protoplanetary discs feeding into them.
Thus, constraining the requirements for planetesimal formation in a given scenario, and comparing to dust evolution models is one way to make progress in understanding which mechanisms succeed and are dominant.

However, the most studied context for a dust-gas streaming scenario in a protoplanetary disc only considers a single size of dust particle, with a single value of the stopping time.
This monodisperse Streaming Instability (mSI) can be viewed as a special case of a more general and physical scenario which accounts for a wide, continuous range of dust sizes, the Polydispserse Streaming Instability (PSI).
Although a number of works have considered multiple dust sizes \citep{2010ApJ...722.1437B,2018A&A...618A..75S} and even approached the continuous limit with a series of multi-species calculations \citep{2019ApJ...878L..30K,2020arXiv200801119Z} the PSI is still largely unexplored.

This paper, the third in the series, employs the tools developed in \citet{paper2} (hereafter \citetalias{paper2}) to study the linear phase of PSI, the fundamentals of which were described in \citet{paper1} (hereafter \citetalias{paper1}).
We focus on extending the range of dust stopping times, and generalizing the dust size distributions to a class of distributions which are plausible outcomes of the dust processing occurring in protoplanetary discs.
These generalisations allow us to draw conclusions about when the linear PSI is conducive to planet formation.

The generalisation of classical mSI (monodispserse Streaming Instability) to PSI allows the consideration of realistic dust size distributions in the protoplanetary disc.
These distributions are non-trivial, as they are the results of the slow processes of dust coagulation-fragmentation and have a dependence on the dust material properties and the disc gas flow
\citep{2011A&A...525A..11B,
2015ApJ...813L..14B, 2016ApJ...818..200E, 2019ApJ...874...26S, 2020SoSyR..54..187K}.

We also consider here in a limited manner the effects of turbulence through a diffusive-viscous model for turbulence, extending the methods used by \citet{2020ApJ...895....4U} and \citet{2020ApJ...891..132C} to a polydisperse context.

\subsection{Review of previous works}
In \citetalias{paper1} we introduced the PSI, and presented a terminal velocity (TV) analysis of the instability.
Our most important finding in this model for well-coupled dust was that for wide dust distributions the instability depends on the size resonance, a resonance related to that found in the Resonant Drag Instability (RDI) theory for the mSI. In RDI theory, fastest mSI growth rates occur when the drift velocity of the dust matches the phase velocity of a gas wave. When considering a size distribution, with a size-dependent drift speed, it may be that a single size in the distribution satisfies an RDI condition, despite the average dust drift being different. We found in \citetalias{paper1} that this size resonance plays a major role in determining PSI growth rates. We finally tested the TV-PSI approximation against the full PSI equations with a direct solution method.

In \citetalias{paper2} we introduced the publically available {\tt psitools} package \citep{psitools} and its advanced numerical methods for solving the PSI linear stability problem.
The direct solver employed in \citetalias{paper1} directly discretizes the entire eigenproblem in size space and solves a matrix form for eigenvalues and eigenvectors.
This direct solver, though faster converging for fast growing modes than the multifluid discretization employed in \citet{2019ApJ...878L..30K} and \citet{2020arXiv200801119Z}, shares the difficulty in attaining results with small relative truncation error and floating point error where the PSI growth rate is small ($\lesssim 10^{-4}\Omega$, where $\Omega$ is the local Keplerian angular velocity).
The root finding algorithm introduced in \citetalias{paper2} uses a scalar formulation of the PSI dispersion relation to find roots in the complex plane. 
This allows the use of high accuracy numerical quadrature schemes for integrals over dust size space yielding high accuracy roots of the dispersion relation, and hence pushing the truncation error down to the floating point error limit on the computation of the dispersion relation.
This yields dramatically more accurate growth rates while also dramatically reducing computational cost.
We exploit these tools in this paper to survey the most interesting conditions for PSI growth which could potentially result in the formation of planetesimals.

\citet{2019ApJ...878L..30K} first studied the linear evolution of PSI through multi-fluid SI calculations. We have demonstrated agreement with the results of their calculations in \citetalias{paper1} and \citetalias{paper2}. 
Their work found that PSI growth is very slow when the local dust to gas mass ratio is less than unity, and did importantly anticipate that dust evolution may have an important effect on the viability of PSI as a planet formation mechanism.
Recently, \citet{2020arXiv200801119Z} presented numerical calculations of specific cases of multi-fluid SI which display a trend in agreement with our conclusions in \citetalias{paper1} about the role of a critical $\mu$ in enabling growth by the size resonance.
The difference between the size resonance and the resonant drag instability also shows why the PSI behaves in such a different way, with the transition from a monodisperse to a wide dust distribution sometimes increasing, and sometimes decreasing, the instability growth rate.
The sharp boundary of the island of fast growing modes \citet{2020arXiv200801119Z} infer in wavenumber space is rendered in high fidelity by the calculations in this series of papers.
The mapping of the complex valued dispersion relation in \citetalias{paper2} showed how this comes about:  in large regions of wavenumber space growing roots flatten towards the real axis asymptotically as the dust distribution is widened.

\subsection{Dust evolution in protoplanetary discs}
Typical dust distributions in the interstellar medium can be roughly described as a power-law in dust size from $\sim0.01\ {\rm \mu m}$ to $\sim0.2\ {\rm \mu m}$ \citep{2011piim.book.....D}.
Dust grains likely grow significantly through coagulation during protostellar collapse, significantly depleting the population of the smallest grains \citep{2020arXiv200704048G}.
Once in the circumstellar disc environment, dust evolution is primarily driven by collisions. The outcome of these collisions depends on the sizes of dust involved, the structure of the aggregates, and the energy of the collision.
For the smallest grains, the dust mutual relative velocities are dominated by Brownian motion, and the gentle collisions result in sticking leading to the growth of aggregates.
Once they have grown larger, the aggregate grains enter a regime where their mutuaal velocities are driven mainly by turbulent motions of the disc gas, which results in faster collisions.
At a critical scale these velocities grow fast enough for the collisions of like-sized grains to result in fragmentation, limiting the growth of the aggregates \citep{2018haex.bookE.136A}.
In the inner parts of a disc, the gas headwind on large aggregates may alone be enough to tear parts off, limiting their growth without particle-particle interactions \citep{2020MNRAS.496.4827R,2020ApJ...898L..13G},
Dust collisions at intermediate scales and unequal-size collisions have further effects, resulting in bouncing, compaction, erosion, or other processes \citep{2010A&A...513A..56G}.
A single dust population does not spend the lifetime of the disc merely interacting locally. Dust of differing sizes and physical properties drifts radially and are lofted vertically to different extents.
Structures in the disc, such as azimuthal pressure bumps, gap and dead zone edges, and vortices likely play a role in the evolution of the dust population, retaining dust which would otherwise drift radially into the star.
This complicated landscape of processes evolves the dust distribution present at any given location of the disc, in a way which is not yet well understood, but must result in the fuel for planet formation.
Understanding the PSI and how the growth rates correspond to dust distributions will reveal which dust evolution scenarios are conducive to planet formation.

Observation of discs can detect the scattered light and/or thermal emission from dust from $0.01\ {\rm\mu m}$ to cm sizes.
The observational signature of dust is a complicated convolution of the physical condition and the radiative transfer properties  (such as the opacity, albedo, and polarization) of the entire population of dust.
These radiative transfer properties can have particularly detailed dependence on the physical morphology and mineralogical properties of the dust grains.
This difficulty currently limits what is observationally known about the dust size distribution in discs.
Beyond thermal continuum emission, scattered light observations and details of the system's spectral energy distributions in the optical and near-infrared indicate the presence of dust particles with sizes $\gtrsim 10\ {\rm \mu m}$ high in the disc atmosphere
\citep{2020arXiv200105007A}.
Protoplanetary disc observations show large populations of grains even in old discs, where simple models would suggest the dust should have been depleted by drifting onto the star,
This seeming contradiction indicates a possible role of azimuthal pressure bumps, trapping and slowing dust on its radial journey
\citep{2018ApJ...869L..41A,2018ApJ...869L..46D}.

\subsection{Where might mSI form planetesimals?}
The proposed role of SI-class instabilities in planet formation of producing planetesimals relies on the instability gathering dust overdensities in the disc which exceed the local Roche density 
at large scales, and turbulent diffusion at small scales, allowing them to collapse into self-gravitating planetesimals \citep{2007Natur.448.1022J,
    2020ApJ...895...91G,2020arXiv200710696K}.
 We study in this work only linear instability growth, but the full scenario of planetesimal formation follows from growth to nonlinear amplitudes leading to gravitational collapse of these dust clouds.
To aim our investigation at a useful parameter regime, we have examined existing evidence about the regime where SI driven planetesimal formation is likely to occur, and judge form the current literature what parameters may be optimal.

Simulations producing streaming instability and gravitational collapse for monodispserse particles typically employ particles with stopping times $\sim 1 \ \Omega^{-1}$. Notable examples are 
\citet{2007Natur.448.1022J} which used particles with $\taus=0.25$--$1\ \Omega^{-1}$ for the first demonstration of SI driven gravitational fragmentation, and 
\citet{2019NatAs...3..808N} which modelled the formation of trans-Neptunian objects employing particles with $\taus=0.3$--$3\ \Omega^{-1}$.

However, due to the difficulty of producing such large particles in coagulation-fragmentation models of dust, another research theme has been an effort to probe the lower limit of particle stopping time required.
This minimum dust stopping time criteria for mSI growth of dust clumps in full simulations is a function of the minimum dust-gas ratio \citep{2015A&A...579A..43C}, but 
\citet{2017A&A...606A..80Y} suggested in any case the lower limit of mSI viability for forming dust-rich filaments in stratified simulations is  $\taus=10^{-2}$--$10^{-3}\ \Omega^{-1}$.
These latter studies probed stratified nonlinear streaming instability, but did not include gravitational collapse.

In three dimensional stratified simulations, which are able to include particle settling and lofting, 
the fundamental parameter is not a local gas/dust volume density ratio, but instead the surface density ratio, and the local dust to gas ratio is a result largely of dust settling to the midplane.
The results of \citet{2017A&A...606A..80Y} show particles with $\taus=10^{-3}\ \Omega^{-1}$ settling to a midplane layer with dust to gas ratio order unity before forming SI filaments.
The collapse to planetesimals successfully from mSI clumps with $\taus=6\times10^{-3}\ \Omega^{-1}$ was shown in \citet{2017ApJ...847L..12S}, and 
 for a range of disc gas radial pressure gradients with $\taus=5\times10^{-2}\ \Omega^{-1}$ \citep{2019ApJ...883..192A}.
\citet{2020arXiv200801727C} found a tentative requirement of particles with at least $\taus=0.2\ \Omega^{-1}$ for mSI driven fragmentation to occur in 3D simulations including the effect of a background long-lived azimuthal pressure bump, which enhances the concentration of particles.
However, smaller $\taus\simeq 0.02\ \Omega^{-1}$ particles did not drive the formation of fragments, and the dust to gas  density ratio only rarely exceeded unity in the simulation volume.\footnote{In \citet{2020arXiv200801727C} the gas density enhancement in the pressure bump does cause the $\taus$ of the particle to vary by location in the domain, but by less than the rough order of magnitude in this discussion.}
Separately, a meta-analysis of previous three dimensional nonlinear simulation results by \citet{2020ApJ...895....4U} suggested that local volumetric gas to dust mass ratios in the midplane exceeding unity, and dust larger than $\taus \gtrsim 0.1\ \Omega^{-1}$ are required to produce mSI driven fragmentation.
\citet{2020ApJ...895....4U} suggests that this $\taus \gtrsim 0.1\ \Omega^{-1}$  criteria comes from the effect of turbulence in the disc gas, beyond the midplane turbulence created purely by a single small particle size streaming through the gas.

Nonlinear SI drives turbulence through gas dust interaction, and ought to occur in a 
background with turbulence driven by other instabilities at some level.
\citet{2020ApJ...895....4U} include a model for turbulence in their linear calculations and find only a very small region of viable parameter space for mSI. 
Attempting to include the effects of turbulence in a similar formalism,
 \citet{2020ApJ...891..132C} suggest a minimum dust stopping time as a function of Shakura-Sunyaev turbulence parameter $\alpha$ of $\taus \gtrsim \alpha^{2/3}\ \Omega^{-1}$ is required for appreciable growth of the linear mSI instability.
For $\alpha < 10^{-2}$ this is a looser stopping time criteria than the one for the formation of fragments from the simulations of \citet{2020arXiv200801727C}.

In summary, the current understanding of nonlinear mSI suggests particle stopping times on the order of $\taus \sim 0.1\ \Omega^{-1}$ are optimal for the production of planetesimals, although under certain assumptions much larger and smaller stopping times may suffice.
This value suggests a centre for the parameter range for the peak particle sizes we will examine in linear PSI and figure-of-merit on which to evaluate the growth rate of PSI against mSI in the context of planetesimal formation.

\subsection{This paper}
In this paper we present circumstances where PSI shows fast growth, and attempt to find regions where PSI is as robust as mSI.
The possible range of parameters is vast, so we present specific cases which, at the current time, seem most relevant and interesting.
In addition, the depth of existing work on streaming and resonant drag instabilities involving the coupled interaction of dust and gas raises myriad questions of the connections between PSI and its generalisations and these previous works.
There is no way for this survey to be exhaustive, so we have made the underlying code publicly available for the community to analyse other scenarios as they arise.
In this paper, Section~\ref{sec:methods} describes the methods used for our calculations,
Section~\ref{sec:dustparam} presents the models for the dust distributions used and the logic behind them,
Section~\ref{sec:laminar} contains results from calculations of the PSI in a laminar background,
Section~\ref{sec:turbulence} presents results for the PSI including a model for the diffusive effects of turbulence,
Section~\ref{sec:discussion} discusses the results in the context of planet formation, 
and finally conclusions are drawn in Section~\ref{sec:conclusions}.

\section{Methods}
\label{sec:methods}

In this work we employ the methods of the {\tt psitools} package described in \citetalias{paper2} available through the Zenodo.org archive service \citep{psitools}.
In brief, three tools are used here.
First, a direct discretization of the system of equations for the PSI linear stability problem produces approximate eigenvalues and the associated eigenvectors ({\tt psitools.direct}).
Second, a reduction of the PSI linear stability problem to a single dispersion relation equation allows  the application of a complex valued root finder, producing high accuracy eigenvalues ({\tt psitools.psi\_mode}).
Third, a parallel implementation of a grid-refinement method for mapping wavenumber space with the root finder produces maps of the fastest growing PSI mode ({\tt psitools.psi\_grid\_refine}).
Detailed discussion of these methods can be found in \citetalias{paper2}, where also the governing equations in the shearing box are given. 
We adopt only $\eta=0.05\ r_0\Omega^2$, where $r_0$ is the fiducial orbital radius of the shearing box, as the value of the radial pressure gradient in the gas disc in this work, although values vary locally around azimuthal pressure bumps, and globally, in a protoplanetary disc. It will also vary in other PSI contexts such as a protolunar disc \citep{2020LPI....51.2976A}.
Smaller pressure gradient values are known to facilitate particle concentration in mSI \citep{2010ApJ...722L.220B}, so situations where these occur may produce  differing results.

We calculate growth rate maps covering $K_{x,z}\in [10^{-1},10^3]$ where the wavenumbers are nondimensionalized as ${\bf K} = {\bf k}\eta/\Omega^2$ \citepalias{paper2}. This corresponds to wavelengths of $1.4$ to $1.4\times10^{-4}$ pressure scale heights horizontally and vertically in an MMSN at $1~\mathrm{au}$.
At the short wavelength, high wavenumber end of this scale, a minimal turbulent viscous effect will damp growth \citep[][also see Section~\ref{sec:turbulence}]{2020ApJ...895....4U,2020ApJ...891..132C}. 
At low wavenumbers and long wavelengths, the local shearing sheet approximation employed in formulating the linear problem \citepalias{paper1, paper2} has inherent limitations on applicability at scales larger than a pressure scale height when simply considering hydrodynamics \citep{2004A&A...427..855U,2015ApJ...811..121M}.
When in addition considering dust-gas dynamics, the assumption of a constant $\eta$ \citepalias{paper1} may break down at radial scales larger then a pressure scale height, particularly in the presence of an azimuthal pressure bump.
However, for the most part, the fastest growth we find in each map when considering laminar flow is in the high wavenumber (short wavelength) side of the wavenumber space considered.

Finally, we only solve for eigenvalues with growth rates above $2\times10^{-7}\Omega$, due the the technical and computational difficulty of finding roots with these small, physically less important growth rates. 
Further discussion of the structure of the PSI dispersion relation and the associated difficulties can be found in \citetalias{paper2}.

\section{Relevant dust parameters}
\label{sec:dustparam}

Notwithstanding the differences between mSI and PSI, we will focus the survey in this paper on conditions approaching those where mSI has previously been proposed to function well, as the dust and gas disc processes leading the system to the point of instability should be the same.
Indeed, the concept behind choosing the particle size in mSI studies has been that this single particle size is representative of, or the dominant one in, the entire dust distribution.
Hence, the stopping time and dust to gas mass ratio regime centres here on $\taus = 10^{-1}\ \Omega^{-1}$ and $\mu \gtrsim 1$.

As established in \citet{2019ApJ...878L..30K}, \citetalias{paper1}, and \citet{2020arXiv200801119Z}, the PSI is likely most
viable as a planetesimal formation mechanism at dust to gas mass ratios $\mu>1$.
Indeed mSI is also most viable as a dense clump-forming mechanism in this high-$\mu$ regime
\citep{2009ApJ...704L..75J,2020arXiv200110000G,2020MNRAS.tmp.2397S,2020arXiv200801727C}.
In this work, we will focus on the linear evolution of PSI in terms of a  fixed $\mu$, even when considering the diffusive and viscous effects of turbulence.
Particularly in the polydisperse case, no consistent theory exists
for particle settling outside of the trace dust density ($\mu\ll 1$) regime, so self-consistent study of this regime requires either enhanced theories or nonlinear and vertically extended simulations.
It should however be anticipated that the effect of turbulence, generated by other magnetohydrodynamic or  hydrodynamic instabilities, or the PSI itself can be expected to be a crucial mechanism for setting the criteria which determine when planet formation follows from the PSI.

While in the PSI, the convenient variable for parameterizing the size of dust is the stopping time $\taus$, dust coagulation models usually work in terms of dust size $a$ directly. In the Epstein drag regime, $\taus \propto a$ \citepalias[][eq.~35]{paper2}.
This makes it possible to simply parameterize the dust distributions in terms on $\taus$ when convenient, and perform a trivial change of variables to make the gas-dust drag integral in the gas momentum equation \citepalias[][eq.~38]{paper2} to be over the range in stopping time $taus$ instead of particle size $a$. In this section we describe the dust distributions in terms of the size $a$, but in describing our results for  PSI growth we will use the stopping time $\taus$.

\subsection{Models for evolved dust size distributions}
\label{sec:dustdists}

\begin{figure}
    \centering
    \includegraphics[width=\columnwidth]{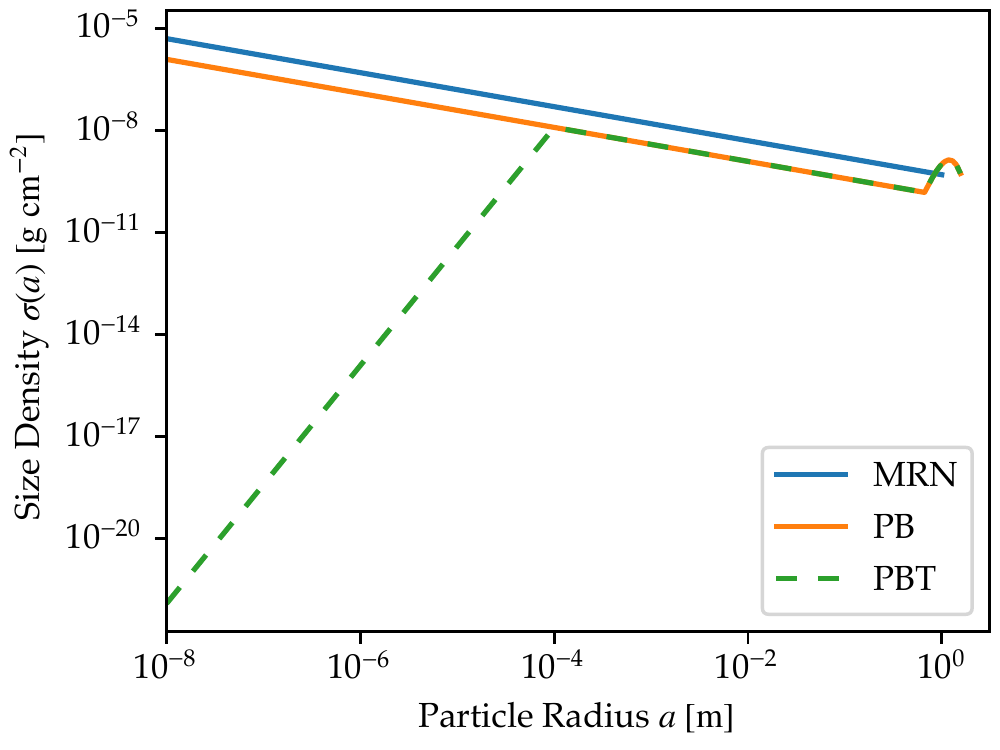}
    \caption{Examples of the MRN, PB, and PBT dust distributions in terms of the size density $\sigma$ as a function of particle radius $a$, for the same total dust volume mass density $10^{-9}\ {\rm[g\ cm^{-3}]}$, and peak particle size $a_{\rm P}=1\ \mathrm{m}$. For the PBT example $a_{\rm BT}=10^{-4}a_{\rm P}$.}
    \label{fig:distributions}
\end{figure}

In this paper we consider three dust size distributions as examples of typical results from dust evolution processes.
Solid dust should exist in discs with a wide range of sizes.
The minimum silicate particle size in the interstellar medium is plausibly a single molecule.
For their dust coagulation model 
\citet{2011A&A...525A..11B} specify a monomer size of $0.025\ \mu{\rm m}$, 
and the inclusion of ISM silicate grains down to at least $0.01\ \mu{\rm m}$ is required to reproduce the optical properties of ISM dust \citep{1984ApJ...285...89D,2011piim.book.....D}. 

Collisional dynamics, with dust collision velocities increasing as the particle size grows, may limit the growth of particles when this velocity scale exceeds the fragmentation velocity of aggregates.
If this fragmentation does not limit the size of dust aggregates, particles with stopping times on the order of an orbital period drift inwards through the disc rapidly compared to dust growth timescales \citep{1977MNRAS.180...57W,2012A&A...539A.148B}. 
Thus, the upper size of the range under consideration for PSI ought to be on the order of unity.
For these reasons, in this survey we choose a fiducial range of dust stopping times as $\taus \in [10^{-8}, 1]\ \Omega^{-1}$ as being reasonably representative for the disc, and vary the upper size limit.

\subsubsection{MRN distribution}
The particle size distribution in the interstellar medium is commonly modelled  in terms of the size density $\sigma$ \citepalias{paper2} as a function of particle radius $a$ as
\begin{align}
\sigma_{\rm MRN}(a) \propto a^{3+\beta}, \ \ \beta=-3.5\ ,
\end{align}
referred to as the MRN distribution \citep{1977ApJ...217..425M,1984ApJ...285...89D}.
Although used as a fit to observations of interstellar dust, this distribution is also mathematically the result of a collisional dust evolution process with pure fragmentation or pure coagulation, and is also a reasonable description of the dust size distribution in a debris disc, and the asteroid belt for this reason \citep{1969JGR....74.2531D,1994Icar..107..117W, 1996Icar..123..450T}.
Hence, this is the basic dust size density distribution that could be expected in the dust inherited form the interstellar medium by a protoplanetary disc.
It has previously been used to study PSI in \citetalias{paper1}, \citetalias{paper2}, and in the works by \citet{2019ApJ...878L..30K} and \citet{2020arXiv200801119Z}.
However, we should expect gas damping of dust velocities in a protoplanetary disc to modify the dust evolution from pure fragmentation.

\subsubsection{Powerlaw-Bump (PB) distribution}
In protoplanetary discs, planet formation can be expected to proceed from dust which has evolved in this context from that found in the interstellar medium.
In the disc, dust relative velocities are smaller then in the ISM, and temperature-density conditions allow for the condensation of further chemical species.
Thus, dust evolution through coagulation-fragmentation allows the formation of  larger dust grains.
Additionally, aerodynamic effects can size-sort dust, resulting in altered dust distributions independent of the coagulation-fragmentation evolution.
As this processing has a very large number of parameters, we do not attempt here to produce a full range of possible dust distributions for all regions of protoplanetary discs.
Instead, we consider a simple model derived from the work of  \citet{2011A&A...525A..11B} and \citet{2015ApJ...813L..14B} which describes the most apparently common outcomes of dust processing in discs from a fragmentation-coagulation model. We caution that here we only attempt to consider the most common and widely used dust evolution, and in general other special cases not anticipated here might become important.
One of the motivations in making the {\tt psitools} package public is to enable the community to analyse linear PSI with specific dust distributions as they arise in the future.

This recipe can be written as of a power-law in particle number density as a function of radius
\begin{align}
f(a) = a^{\beta} \label{eq:fplain}\, ,
\end{align}
and a  Gaussian bump,
\begin{align}
b(a) = 2.0 f(a_{\rm L}) \exp\left( \frac{-\left(a-a_{\rm P}\right)^2}{w}\right)\, ,
\end{align}
with the width
\begin{align}
w = \max\left(
        \frac{\min(|a_{\rm R}-a_{\rm P}|,|a_{\rm L}-a_{\rm P}|)}{\sqrt{\ln(2)}}, (0.1)a_{\rm P}
        \right)\, ,
\end{align}
and then
\begin{align}
F(a) = \left\{\begin{array}{cll}
f(a) & \text{if} & a \leq a_{\rm L}\\
\max\left(f(a),b(a)\right) & \text{if} & a_{\rm L} \leq a \leq a_{\rm P} \\
b(a) & \text{if} & a_{\rm P} \leq a \leq a_{\rm R}\\
0 &\text{else.}& 
\end{array}
\right.
\end{align}
Which then, normalized, gives the dust size density as
\begin{align}
\sigma_{\rm PB}(a) = a^3 F(a)\left/\int_{a_{\rm min}}^{a_{\rm max}}  w^3 F(w)\, {\rm d}w\right. . \label{eq:PB}
\end{align}
We refer to this distribution as PB (Powerlaw-Bump).
Physically, \citet{2011A&A...525A..11B} prescribe the Gaussian bump at the top end of the distribution as the result of a local change in size space to the coagulation-fragmentation fluxes due to the prevalence of cratering collisions between small and large dust grains.
For simplicity, we consider here specific values of $a_{\rm L}$ and $a_{\rm R}$ which the \citet{2011A&A...525A..11B} recipe generates for disc conditions typical of a minimum mass solar nebula at 1--10~au (Appendix~\ref{app:bump}):
\begin{align}
a_{\rm L}&= \frac{2}{3} a_{\rm P}\, , \label{eq:aL} \\
a_{\rm R}&= 1.56 a_{\rm P}\, . \label{eq:aR} 
\end{align}
In addition we adopt $\beta=-3.5$ for the power-law component as this fiducial power-law index is close to the simplified value adopted in \citet{2015ApJ...813L..14B}, and facilitates the comparison of this density distribution and the MRN distribution.

\subsubsection{Powerlaw-Bump-Tail (PBT) distribution}
Below the critical dust size where Brownian motion dominates over turbulence as a driver of dust collisions, the \citet{2011A&A...525A..11B}  model includes a steeper power-law regime of the dust distribution.
To include this steep tail between the dust monomer size $a_0$ and the  Brownian motion-turbulence transitions scale $a_{\rm BT}$ we follow the form from \citet{2015ApJ...813L..14B} 
and substitute $f(a)$ in the PB distribution with $f_{\rm tail}(a)$ defined as:
\begin{align}
f_{\rm tail}(a) = 
\left\{\begin{array}{cll}
a_{\rm BT}^{\beta} \left(\frac{a}{a_{\rm BT}} \right)^{1/2} &\text{if} & a<a_{\rm BT} \\
 a^{\beta} &\text{else.} & \ 
\end{array}
\right. \label{eq:ftail}
\end{align}
We refer to the dust size density distribution Equation~(\ref{eq:PB}) with this tail form Equation~(\ref{eq:ftail}) substituted for Equation~(\ref{eq:fplain}) as the Powerlaw-Bump-Tail (PBT) distribution.

Examples of the three distributions used in this work are shown in Figure~\ref{fig:distributions}. Note all the distributions are what is commonly referred to as `top-heavy' in that by particle mass, the total dust density is dominated by the largest particles. In Figure~\ref{fig:distributions} this is not visually obvious as the size density $\sigma(a)$ is parameterized by particle radius $a$ not particle mass.

\subsection{Model for Gas Turbulence}
\label{sec:turbmodel}

To explore the possible impact of turbulent mixing on the linear growth of PSI, we use a turbulence model to parametrise the effect of turbulent mixing on the gas and dust fractions. Turbulence models usually model subgrid turbulence as a viscous term in the fluid, as turbulence drives momentum diffusion. Analogously, turbulent eddies can move around dust particles suspended in the fluid, driving dust diffusion. Gas turbulence and the consequent dust diffusion can be described consistently using a viscous turbulence model for the gas and a mass diffusion for the dust.

We follow the formulation of \citet{2020ApJ...891..132C}, which uses $\alpha$-viscosity as the viscous term in the gas momentum equation, and a corresponding diffusive term in the dust continuity equation, extending it to our polydisperse dust equations. These terms are an extra turbulent diffusion of momentum, adding to the gas momentum equation \citepalias{paper2}
\begin{align}
 \partial_t\vg + (\vg\cdot\nabla)\vg =& -\frac{\nabla P}{\rhog} + \mathbf{F_{\mathrm{drag}}} +\frac{1}{\rhog}\nabla \cdot \mathsf{T}_{\mathrm{visc}}\,
 ,\label{eq:govsecondturb}
\end{align}
where the final term is the added viscous diffusion term with stress tensor
\begin{align}
\mathsf{T}_{\mathrm{visc}}=\rhog \nu \left[\nabla \vg + \nabla \vg ^{\dagger} - \frac{2}{3}\mathbb{I}\nabla\cdot\vg\right].
\end{align}
Here $\mathbb{I}$ is the identity tensor.
The turbulent viscosity $\nu$ is then defined following \citet{1973A&A....24..337S} 
\begin{align}
    \nu=\alpha c^2 \Omega\,,
\end{align}
where $\alpha$ is the dimensionless parameter describing viscosity.
A corresponding turbulent diffusion term is also added to the dust continuity equation \citepalias{paper2}
\begin{align}
\partial_t\sigma + \nabla\cdot(\sigma \ud ) =& \nabla\cdot\left[D(\taus)\rhog\nabla\mu\right]\, ,\label{eq:govthirdturb}
\end{align}
where the final term adds dust mass diffusion, with $ D(\taus)=\delta(\taus) c^2 \Omega $ a diffusion coefficient depending on $\taus$. The term $\delta$ relates the turbulent parameter $\alpha$ to the rate of dust diffusion and the dust stopping time $\taus$ \citep{2007Icar..192..588Y},
\begin{align}
\label{eqn:dust-turbulence-coupling}
    \delta=\frac{1+\taus\Omega+4\left(\taus\Omega\right)^2}{\left(1+(\taus\Omega)^2\right)^2}\alpha,
\end{align}
encoding how larger dust particles are less subject to turbulent diffusion than smaller well coupled particles. The dust diffusion coefficient $D\left(\taus\right)$ is therefore a function of the grain stopping time $\taus$. Small grains have $\taus\ll 1$, making $\delta(\taus)\approxeq\alpha$. However, as grains grow, $\delta(\taus)$ gradually goes to $0$ as the larger grains are less subject to gas driven stirring due to their larger mass to cross-section area ratio. 
This model represents only the averaged local effects of turbulence, and does not capture turbulent clumping and other turbulent fluctuations. The model also does not consider the effect of dust on the gas turbulence.
It also assumes only isotropic turbulent diffusion and viscosity effects, whereas at least one possible driver of turbulence, the Vertical Shear Instability \citep{2013MNRAS.435.2610N} is strongly anisotropic, with much greater vertical mixing than horizontal \citep{2017A&A...599L...6S}. 

\citet{2020ApJ...891..132C} and \citet{2020ApJ...895....4U} have, in the mSI case, directly connected the strength of turbulence in this model to a dust scale height, and used this to map a dust column density to a midplane dust to gas ratio $\mu$.
However, we do not attempt to make such an identification in this work.
The analytical model used for dust lofting in \citet{2020ApJ...891..132C} and \citet{2020ApJ...895....4U} assumes a trace dust density $\mu\ll 1$, whereas planet formation driven by PSI is most viable at $\mu \gtrsim 1$.
The results we show do not depend on a dust lofting model, as it is unclear whether such models are applicable in the regimes studied.
Instead, in this work we specify the dust to gas mass density ratio and turbulence
strength independently, not invoking a model for the vertical
thickness of the dust distribution.
We similarly do not include the slow viscously driven radial
motion of gas in the midplane which should arise as a consequence of momentum diffusion as this brings complex dependence on disc model and location parameters \citep{1984SvA....28...50U,2017ApJ...837..101P}.
Self-consistent models for dust lofting, including polydisperse dust and valid at $\mu\sim 1$, are thus a topic of great future interest, as the column density of dust is a much more natural input parameter to the PSI initial conditions in a planetesimal forming context than the volume density of dust.

\section{PSI growth rates and eigenmode structures - Laminar}
\label{sec:laminar}
\begin{figure*}
	\includegraphics[width=0.9\textwidth]{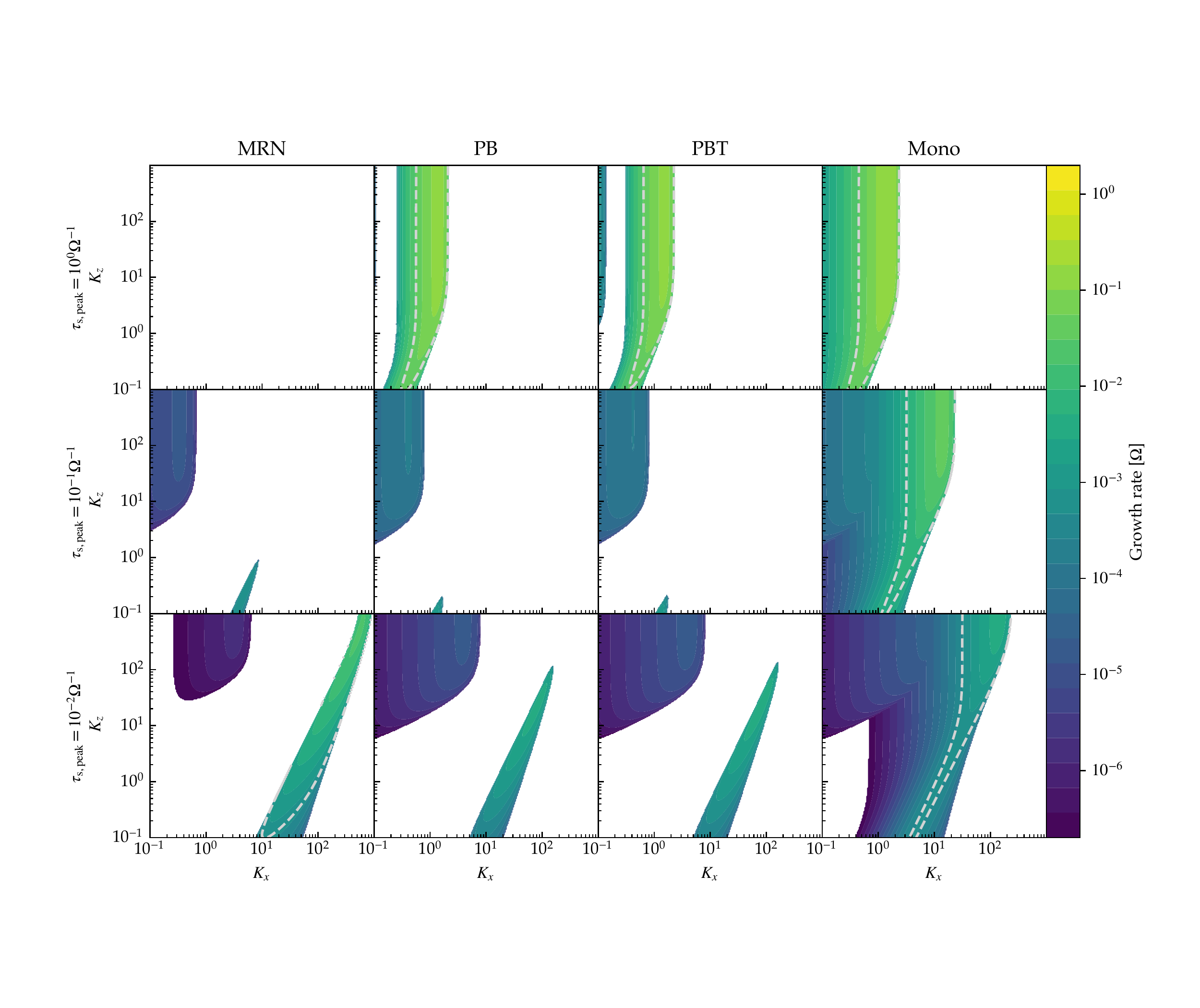}
    \caption{Growth rates (in units of $\Omega$) for $\mu=0.5$ truncated at $2\times10^{-7}\ \Omega$ with MRN, PB, PBT, and monodisperse dust distributions. {\sl Grey dashed contour:} Longest growth timescale exceeding dust drift  time over a length $0.05r_0$.}
   \label{fig:compmu0p5}
\end{figure*}

\begin{figure*}
	\includegraphics[width=0.9\textwidth]{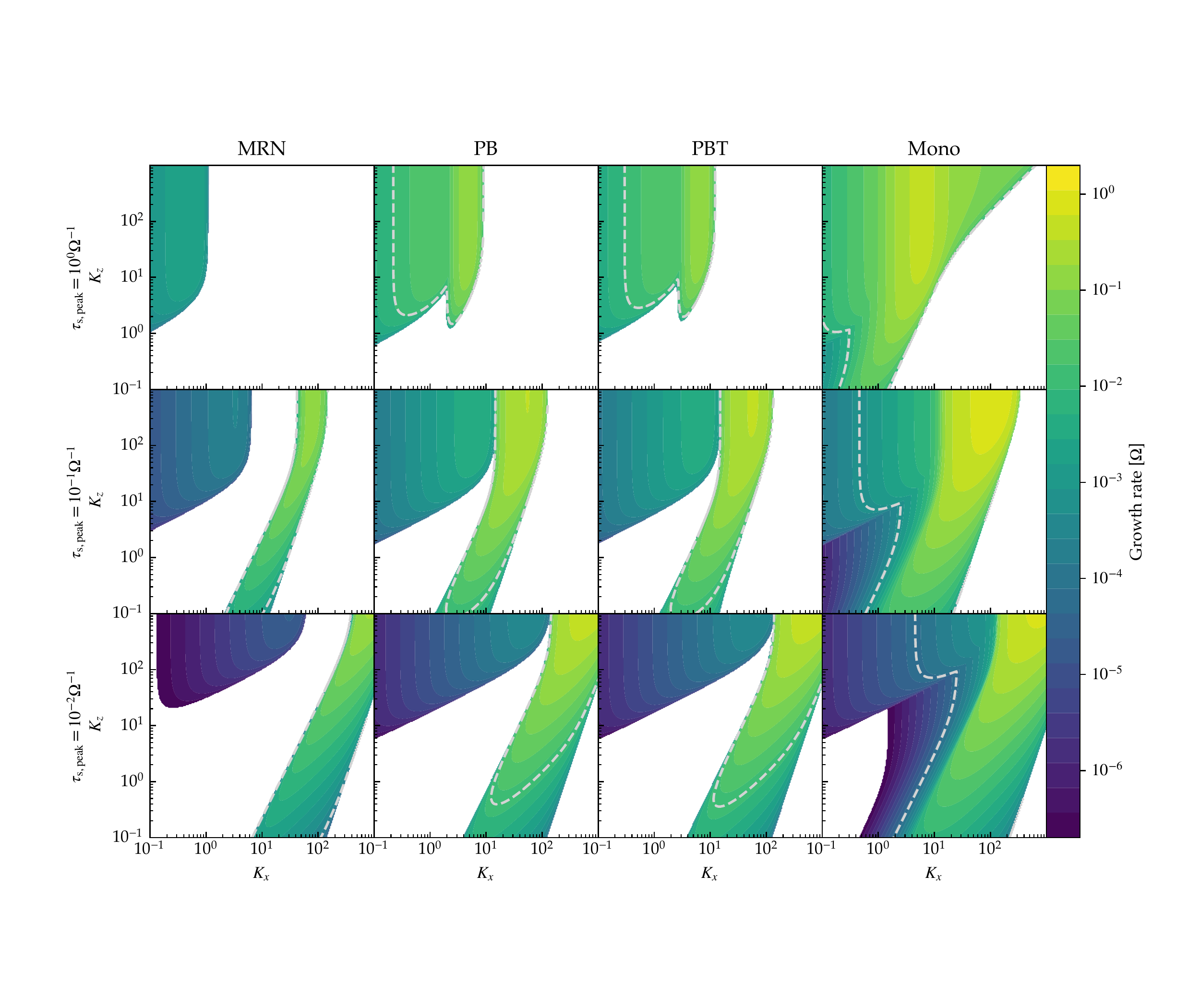}
    \caption{PSI growth rates (in units of $\Omega$) for $\mu=3$ truncated at $2\times10^{-7}\ \Omega$ with MRN, PB, PBT, and monodisperse dust distributions. {\sl Grey dashed contour:} Longest growth timescale exceeding dust drift time over a length $0.05r_0$.}
    \label{fig:compmu3}
\end{figure*}

\begin{figure}
\begin{center}
	\includegraphics[width=\columnwidth]{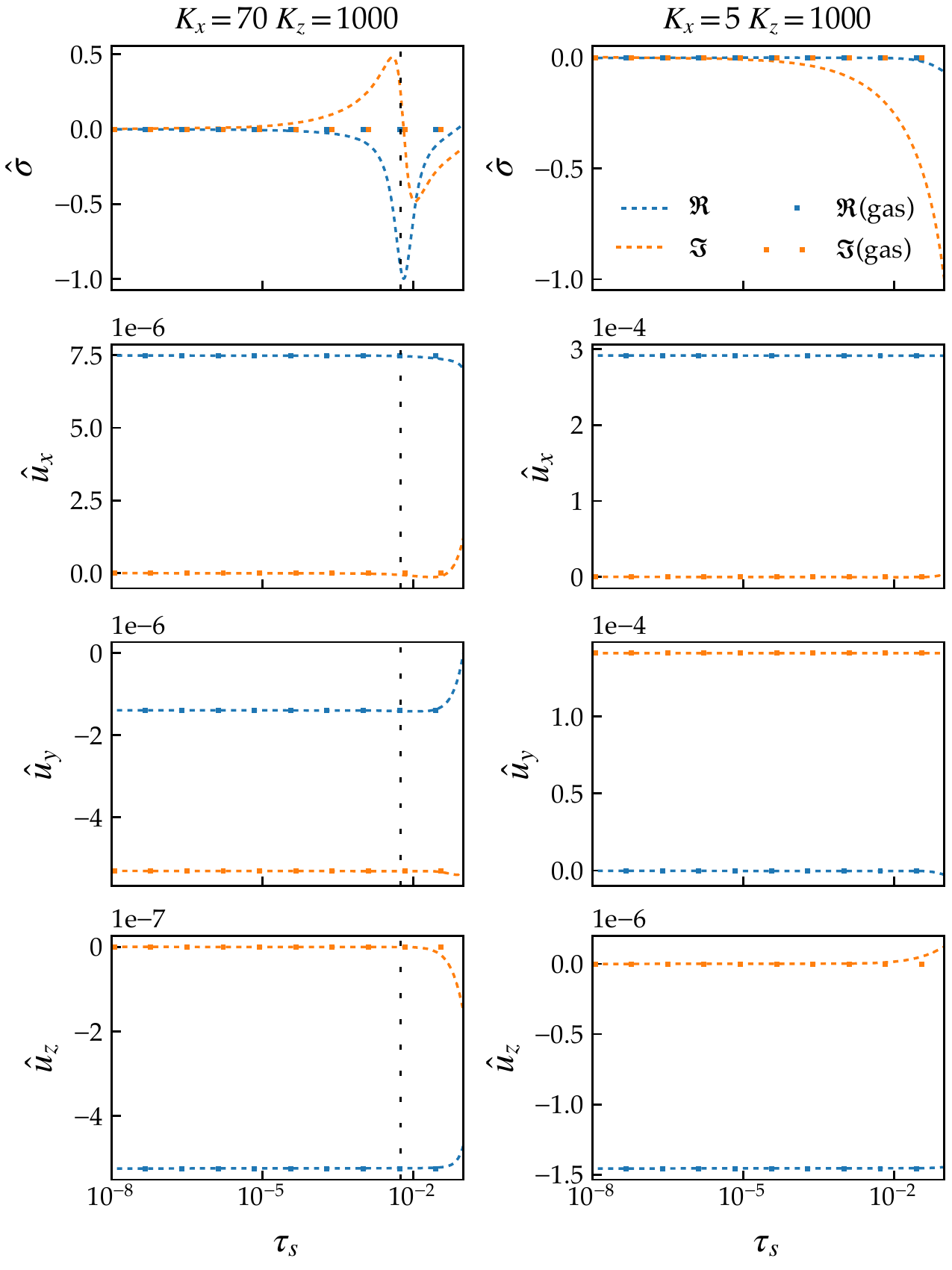}
    \caption{Dust components of eigenfunction for the fastest growing mode with $\mu=3$ and $\tauspeak=10^{-2}\ \Omega^{-1}$. {\sl Left:} $K_x=70$ $K_z=1000$, size resonance present. {\sl Right:} $K_x=5$ $K_z=1000$, no size resonance present. {\sl Vertical dashed lines:} Position where the mode phase velocity matches dust radial drift velocity, giving rise to the size resonance.
    Each panel for a dust component includes dotted lines for the corresponding scalar component of the gas perturbation from the eigenvector.
    Eigenfunctions are normalized to set $\Im(\hatvgx)=0$ and $|\max(\hatsigma)|=1$. }
    \label{fig:mu3eignefuncs}
\end{center}
\end{figure}

\begin{figure*}
	\includegraphics[width=0.9\textwidth]{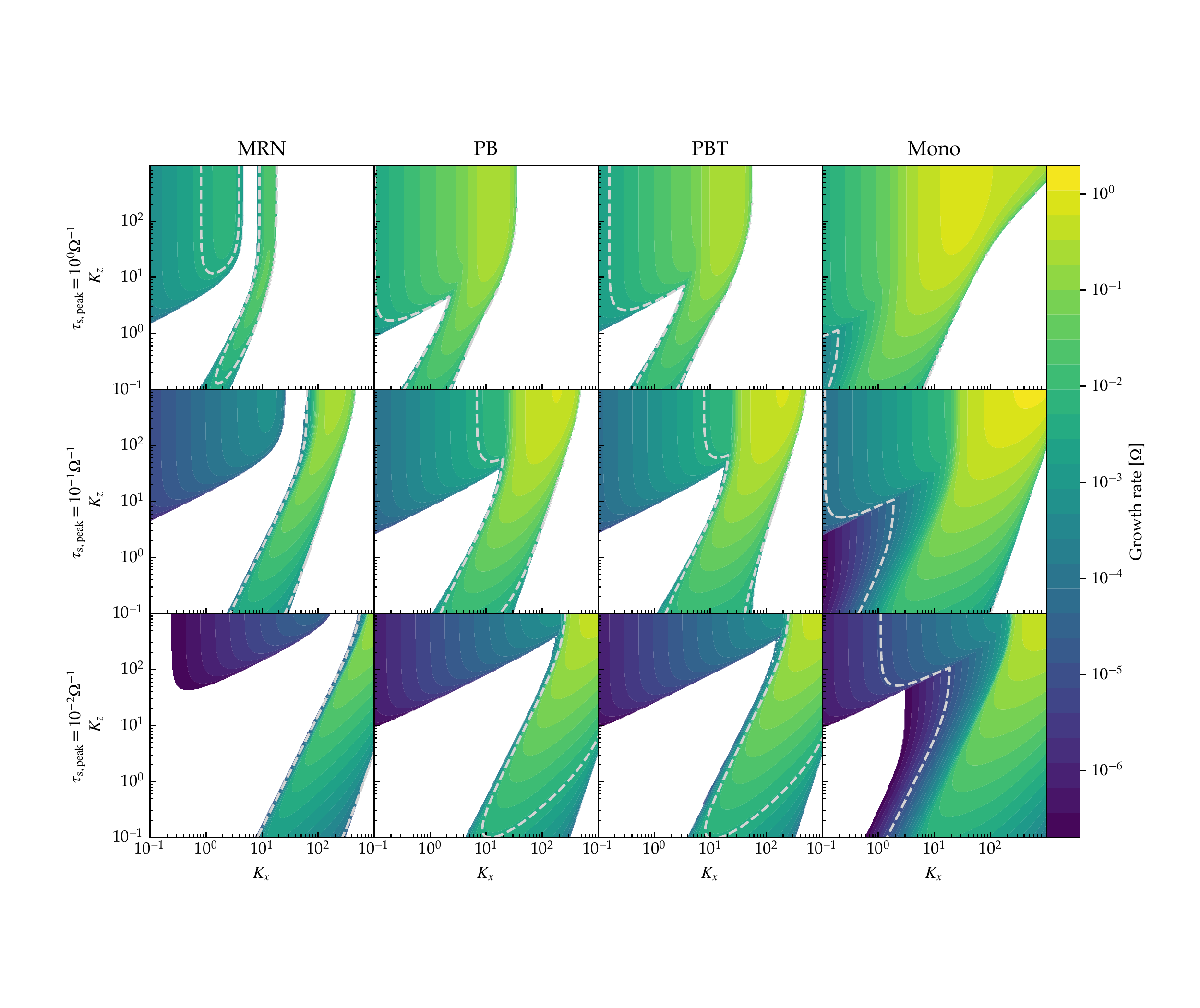}
    \caption{PSI growth rates (in units of $\Omega$) for $\mu=10$, truncated at $2\times10^{-7}\ \Omega$ with MRN, PB, PBT, and monodisperse dust distributions. {\sl Grey dashed contour:} Longest growth timescale exceeding dust drift time over a radial distance $0.05r_0$.}
    \label{fig:compmu10}
\end{figure*}

\begin{figure}
\begin{center}
	\includegraphics[width=0.9\columnwidth]{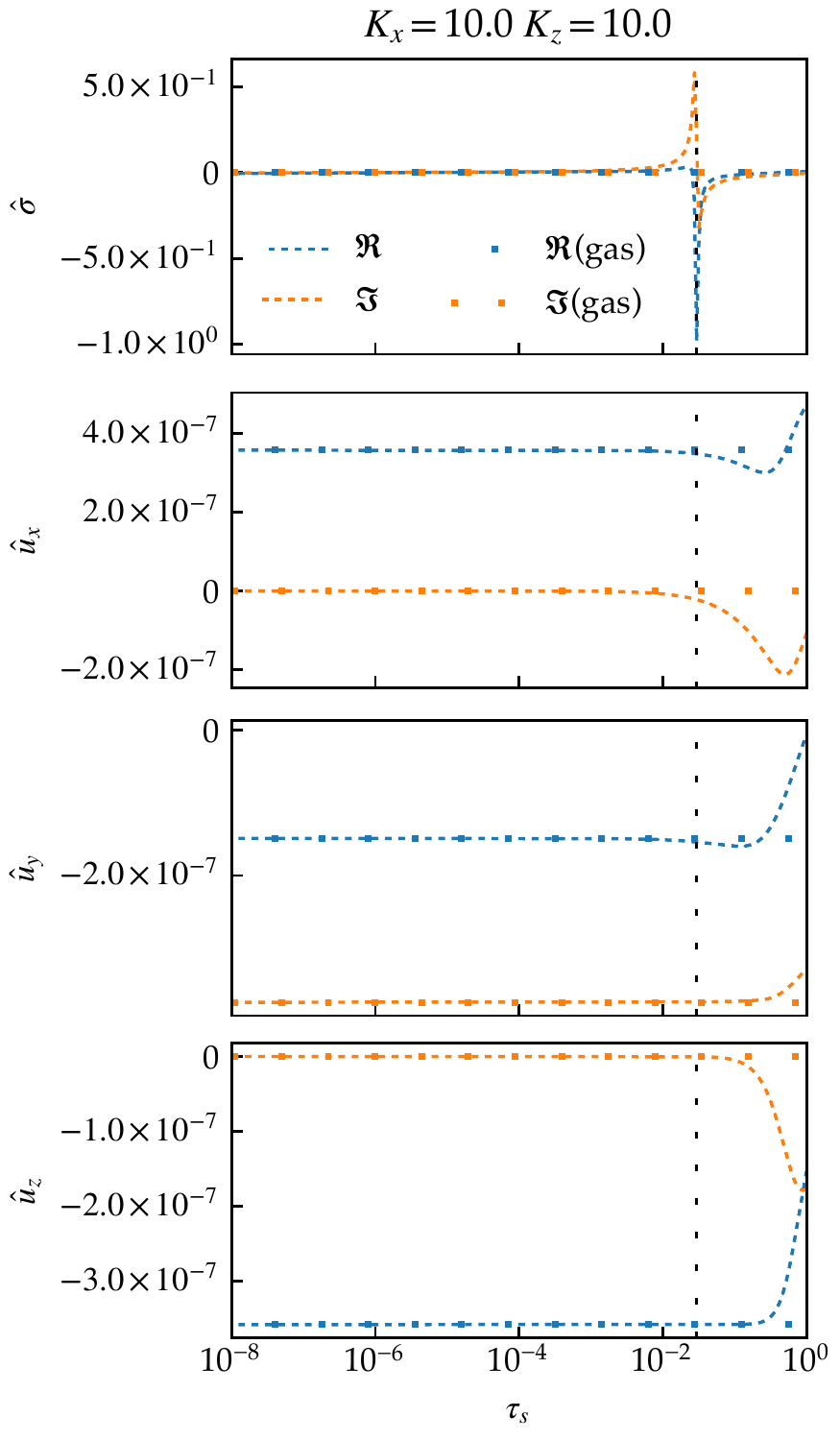}
    \caption{Dust components of eigenfunction for the fastest growing mode with $\mu=10$, $\tauspeak=10^{0}\ \Omega^{-1}$, $K_x=10$, and $K_z=10$. {\sl Vertical dashed lines:} Position where the mode phase velocity matches dust radial drift velocity, giving rise to the size resonance.
    Each panel for a dust component includes dotted lines for the corresponding scalar component of the gas perturbation from the eigenvector.
    Eigenfunctions are normalized to set $\Im(\hatvgx)=0$ and $|\max(\hatsigma)|=1$. }
    \label{fig:MRNmu10taus1eigenfuncs}
\end{center}
\end{figure}

\begin{figure}
\begin{center}
	\includegraphics[width=1.0\columnwidth]{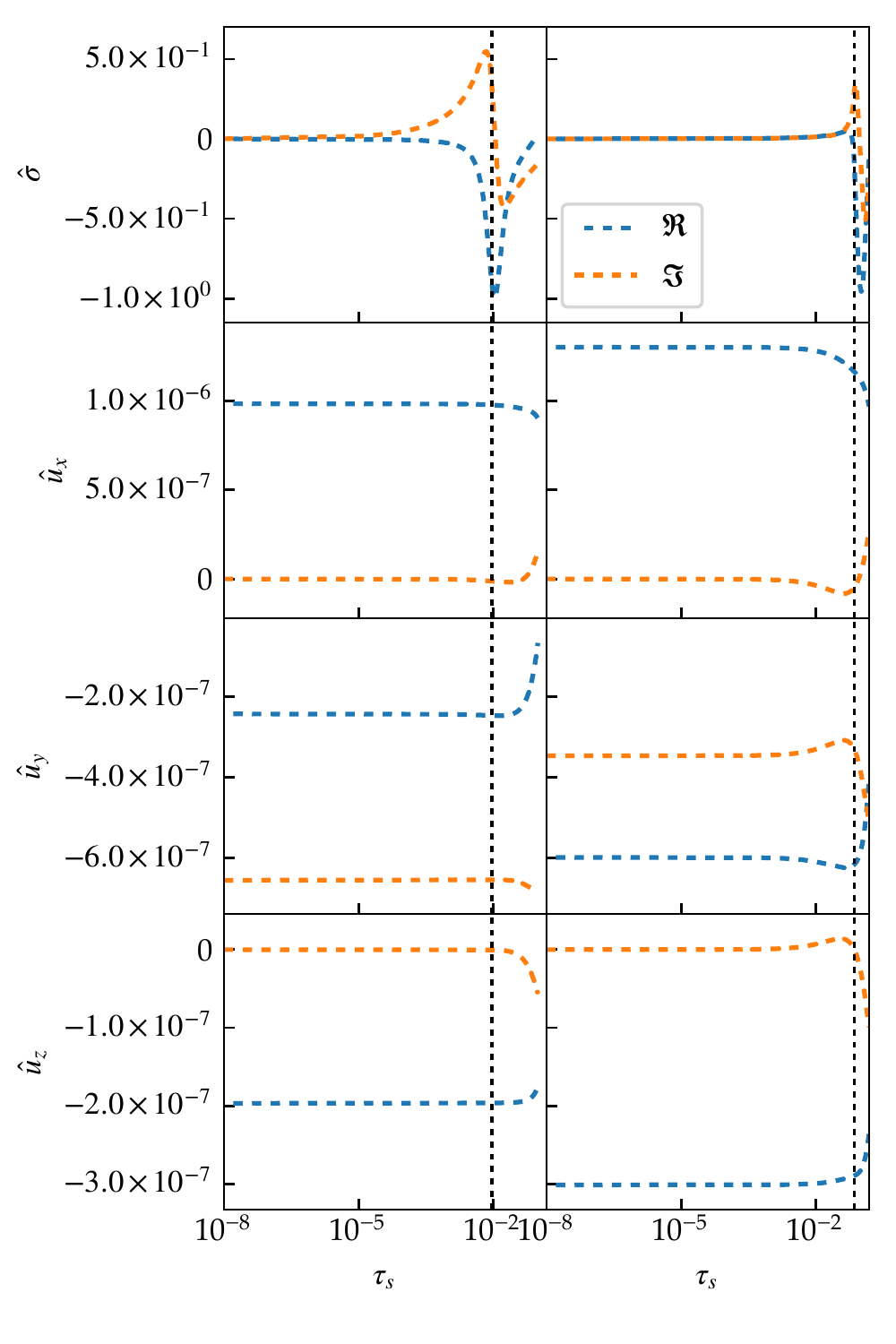}
    \caption{Dust components of eigenfunction for the fastest growing modes with $\tauspeak=10^{-1}\ \Omega^{-1}$, $\mu=10$. {\sl Left:} MRN dust distribution {\sl Right:} PB dust distribution. {\sl Dashed lines:} Position where the mode phase speed matches $u_{{\rm d},x}-\bar{u}_{{\rm d},x}$, giving rise to the size resonance, in the MRN case
    $\taus = 1.2\times10^{-2}\ \Omega^{-1}$ and PB case
    $\taus = 6.0\times10^{-2}\ \Omega^{-1}$.
    Eigenfunctions are normalized to set $\Im(\hatvgx)=0$ and $|\max(\hatsigma)|=1$.}
    \label{fig:mrn_vs_pb_eignenfunctions}
\end{center}
\end{figure}

\begin{figure*}
	\includegraphics[width=0.9\textwidth]{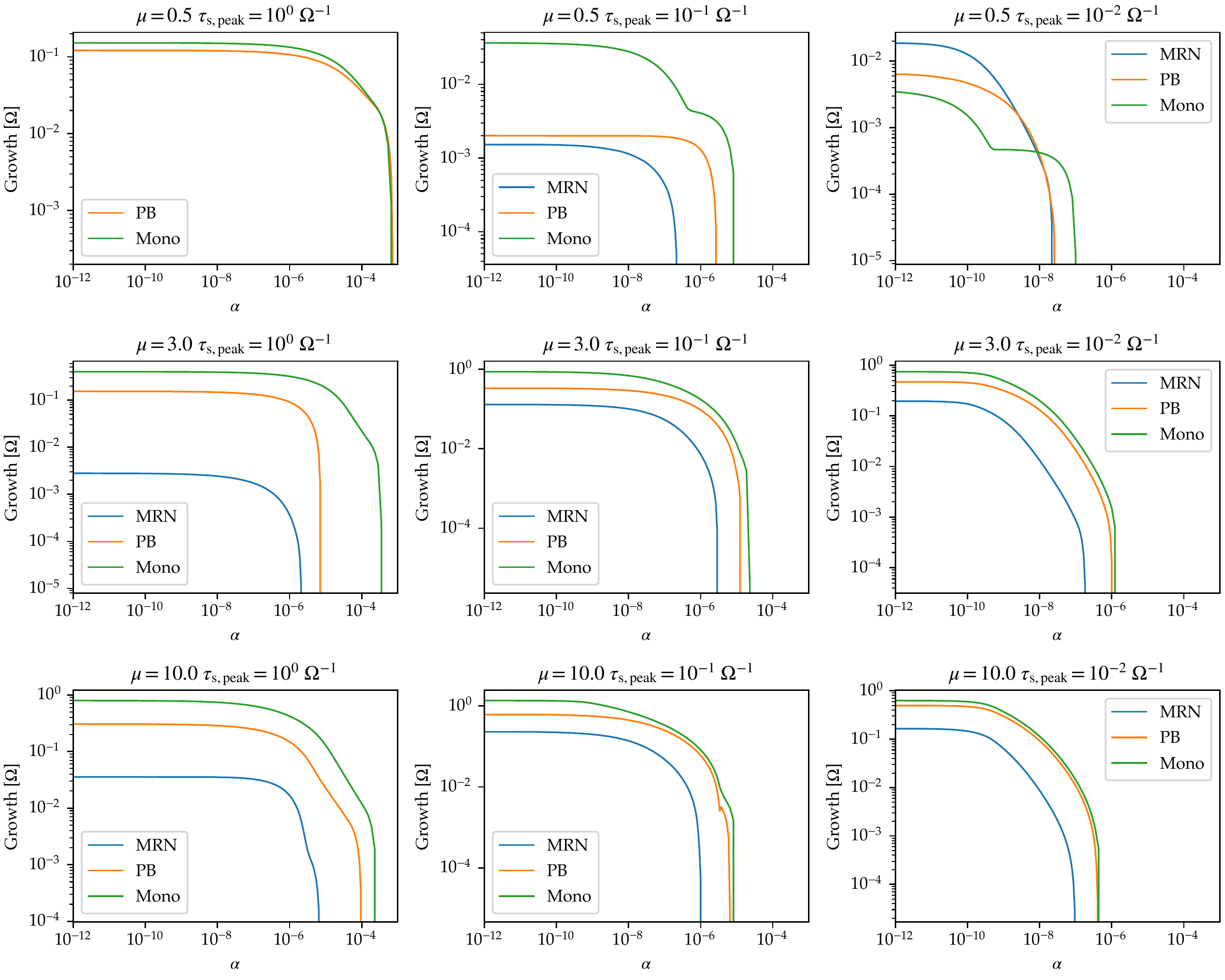}
    \caption{Maximal growth rates (in units of $\Omega^{-1}$) for varying $\alpha$ with MRN, PB, and monodisperse dust distributions at dust to gas ratios $\mu\in {0.5, 3, 10 }$.}
    \label{fig:maximize_alpha}
\end{figure*}

\begin{figure*}
	\includegraphics[width=\textwidth]{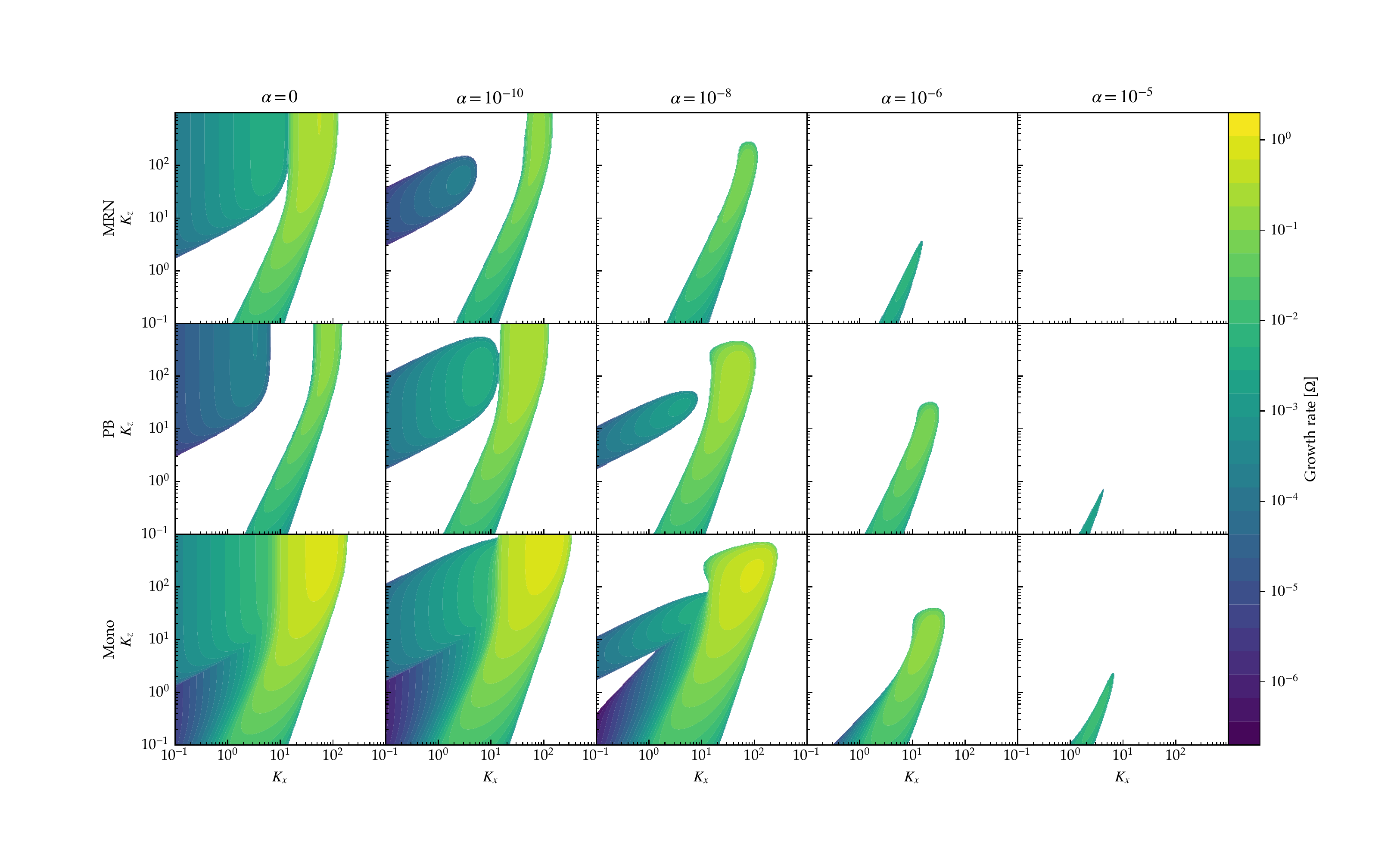}
    \caption{Growth rates (in units of $\Omega$) for $\mu=3$ truncated at $2\times10^{-7}\ \Omega$ varying viscous $\alpha$ for MRN, PB, and monodisperse dust distributions with $\mu=3$ and $\tauspeak=10^{-1}\ \Omega^{-1}$, $\tausmin=10^{-8}\ \Omega^{-1}$ ($\taus=\tauspeak$ for monodisperse).}
    \label{fig:alpha_growth_map}
\end{figure*}

The first set of growth rate results we present is a survey of the instability growth rate mapped over wavenumber, for varying dust size density distribution, varying dust to gas mass ratios $\mu$, and varying peak particle size stopping times $\tauspeak$.
Growth rate maps computed with the grid refinement algorithm with 4 runs per point up to a $241^2$ grid \citepalias{paper2} for $\mu=0.5$ are shown in Figure~\ref{fig:compmu0p5}, for $\mu=3.0$ in Figure~\ref{fig:compmu3}, and $\mu=10$ in Figure~\ref{fig:compmu10}.
In each figure, results for the MRN, PB, PBT, and monodisperse dust distributions are shown.
Each column of the figures corresponds to a value for the peak of the size-density distribution $\tauspeak$, which is not in general the peak of the particle mass distribution.
For the MRN distribution $\tauspeak$ is the maximum dust size (and stopping time), for the PB and PBT distributions it is the peak of the Gaussian bump in these functions, and for the monodisperse dust distribution it is the single dust stopping time present. The three columns show results for $\tauspeak$ of $10^{-2}\ \Omega^{-1}$, $10^{-1}\ \Omega^{-1}$, and  $1\ \Omega^{-1}$. The MRN, PB, and PBT distributions have fixed minimum $\taus$ of $10^{-8}\ \Omega^{-1}$ and for the PBT distribution the transition between power-law regimes is fixed at $\taus_{\rm, BT}=10^{-4}\ \Omega^{-1}$.

In Figures~\ref{fig:compmu0p5},~\ref{fig:compmu3}, and~\ref{fig:compmu10} we include as a figure of merit for the growth rates a contour drawn at the exponential growth time which equals time for the fastest drifting dust to radially draft a distance of $0.05r_0$, which is on the order of the gas pressure scale height. 
This length is also on the order of the largest spatial scales that the local shearing-box approximation used to derive the PSI dispersion relation (\citetalias{paper2}) can describe.

In  Figure~\ref{fig:compmu0p5} the $\mu=0.5$, $\taus=10^0\Omega^{-1}$ cases carry the first surprise.
For the pure power-law MRN distribution, no growth above the $2\times10^{-7}\Omega$ threshold is found in the wavenumber space.
However, when the bump is added at the largest dust sizes in the PB distribution, a feature resembling the growth pattern in the monodisperse case both in wavenumber space and growth rates appears.
This ridge in the PB and PBT cases does also have the important feature of a sharp cut-off on the low-$K_x$ side, in contrast to the monodisperse case.
At the smallest $K_x$ in PB and PBT a second island of more slowly growing modes appears. 
This is similar to the behaviour seen with MRN dust distributions at $\mu>1$, and more analysis of these modes will be presented when discussing those cases.

For the $\mu=0.5$, $\taus=10^{-1}\Omega^{-1}$ cases in Figure~\ref{fig:compmu0p5}, the suggestion of the size resonance is seen in the fastest growing island of modes, but the peak growth rates for all the polydisperse cases are substantially below the monodisperse case.
The left side of the fastest growing island in the MRN distribution island lies at higher $K_x$ then the ridge of fast growth in the monodisperse case, similar to what was demonstrated in the terminal velocity limit in \citetalias{paper1}.

For the $\mu=0.5$, $\taus=10^{-2}\Omega^{-1}$ cases in Figure~\ref{fig:compmu0p5} there is also a surprising contrast between the behaviour for different dust distributions. For these parameters, the maximal growth of PSI is faster than the equivalent mSI case. Interestingly, this is contrary to the assertion made in \citet{2019ApJ...878L..30K} that the MRN-distributed PSI exhibits maximal growth rates smaller than the corresponding mSI, and agrees with the  conclusions of \citet{2020arXiv200801119Z} that such a phenomenon may occur.
This parameter set lies also within the short stopping time regime addressed with the terminal velocity approximation  in \citetalias{paper1} and the results are similar. Note though that in \citetalias{paper1} different wavenumber cutoffs were used, so that the island of slow growth is not visible in Figure 4 of \citetalias{paper1}.
In the MRN case the ridge of fast growth is roughly at higher $K_x$ than in the monodisperse case.
Curiously here, the dust abundance bump at largest dust sizes in the PB and PBT distribution inhibits the growth of PSI, despite this modification making the distribution more central and like the monodisperse case.
Instead of producing the faster size resonance growth of the MRN case, or the secular mSI growth of the monodisperse case, a less optimal middle ground is found.

Most of the growth maps in Figures~\ref{fig:compmu0p5}, \ref{fig:compmu3} and~\ref{fig:compmu10}  contain two regions of growing modes. 
In \citetalias{paper1} is was shown how the size resonance in the terminal velocity limit predicts the onset of growth in a region identified by a cut-off at low $K_x$.
However, a second island of appreciable instability typically appears with a low $K_z$ cutoff, and typically filling all $K_x$ below a cutoff, albeit in some cases a slow decent below the threshold is evident (e.g.~$\mu=0.5$, $\tauspeak=10^{-2}\Omega^{-1}$.
For example, consider the panel for an MRN distribution, $\mu=3$, and $\tauspeak=10^{-1}\ \Omega^{-1}$ in
Figure~\ref{fig:compmu3}. 
Eigenfunctions for two fastest growing modes, one from the size density resonance strip, and one from the low-$K_x$ island are shown in Figure~\ref{fig:mu3eignefuncs}.
The eigenfunctions for $K_x=70$, $K_z=1000$ illustrate the role of the size resonance, with the dust size density having a strong response where the equilibrium radial dust drift velocity matches the radial phase velocity of the mode.
However, in the low-$K_x$ island, no size resonance effect is present, with the dust radial drift velocity and mode radial phase velocity never coinciding. Instead, the strongest response from the dust is for the maximum dust size.
This difference between the two sets of modes, one size-resonant and the other being non size-resonant, will appear again in the examination of the effect of turbulent diffusion on the PSI.

In Figure~\ref{fig:compmu3} the $\mu=3$ runs the PSI growth rates are generally faster than those with $\mu=0.5$.
Particularly for the MRN cases, this reflects the faster growth for the high-$\mu$ PSI, with the $\tauspeak=10^{-2}\Omega^{-1}$ MRN case particularly showing how the size-resonance growth in the terminal velocity limit is shifted to higher $K_x$ than the monodisperse case.
This is also apparent in the $\tauspeak=10^{-1}\Omega^{-1}$ MRN case.
However, in the $\tauspeak=10^{0}\Omega^{-1}$ MRN case this ridge of faster size-resonance growth has disappeared. It should be noted at this long stopping time the terminal velocity approximation does not apply \citepalias{paper1}.

In the $\mu=3$ collection of Figure~\ref{fig:compmu3}, the PB dust distribution gives rise to a hybrid map of fast growth with ridges of fast growth at locations slightly shifted form the monodispserse case, but with sharp cutoffs. In particular, in the  $\tauspeak=10^{0}\Omega^{-1}$ PB case a fast growing ridge is produced where none existed with the MRN distribution, and at approximately the same location as the fastest growing ridge in the monodisperse case.
Again as in the $\mu=0.5$ case, the PBT and PB distributions exhibit very similar growth maps, and the distribution of the smallest dust again has little impact on the end result.

Finally, in  Figure~\ref{fig:compmu10} the $\mu=10$ cases show the behaviour  of PSI at the highest dust to gas ratios in this survey.
Notably, at $\tauspeak=10^{0}\ \Omega^{-1}$ here the MRN case contains a ridge of fast growing modes as seen also at lower $\mu$--$\tauspeak$ combinations. This suggests the size resonance at longer stopping time requires a higher $\mu$ to produce instability.
Inspecting the eigenfunction shown in Figure~\ref{fig:MRNmu10taus1eigenfuncs} one can see how the size resonance occurs here well below the peak of the dust distribution, at $\taus=2.9\times10^{-2}\ \Omega^{-1}$.

The difference between the growth rate maps for the MRN and PB dust distributions leads to the question of how the presence of the bump at the peak of the PB distribution enables faster growth.
Figure~\ref{fig:mrn_vs_pb_eignenfunctions} shows dust eigenmode structures as a function of dust stopping time  for  fastest growing modes in the $\mu=10$, $\tauspeak=10^{-1}\ \Omega^{-1}$ maps for MRN and PB dust distributions as calculated with the direct solver.
These show how the additional abundance of the $\taus\sim 10^{-1}\ \Omega^{-1}$ dust in the PB distribution allows significantly faster growth in the PSI.
In the left column for the MRN distribution, the size resonance and the corresponding fast changes in the dust size density are located at $\taus=9.4\times10^{-3}\ \Omega^{-1}$, whereas with the PB distribution in the right column the resonance is shifted up to $\taus=7.4\times10^{-2}\ \Omega^{-1}$.
This also explains why the short stopping time tail of the size density has little effect on the size resonance growth rates -- this tail contains little mass, and does not change the background dust drift velocities, and so has little effect on the location of the size resonance.

Cross checking of the results from the root finding algorithm by the direct solver for select cases shown here can be found in Appendix~\ref{app:psidirectcomp}.
We have also conducted a further study of dust distributions with very long stopping times in Appendix~\ref{app:longstop}.
Such results are mathematically instructive, but the difficulty of growing dust past $\taus\sim 10\ \Omega^{-1}$ in a coagulation scenario and the lack of physical consistency of the model equations in this regime lessens the relevance of these calculations to planetesimal formation.

\section{PSI growth rates - Turbulent}
\label{sec:turbulence}
To investigate the effect of background turbulence in the disc on the PSI, in this section we employ the diffusive turbulence model terms described in section~\ref{sec:turbmodel}.
Figure~\ref{fig:maximize_alpha} displays the maximal growth rates for the PSI in the wavenumber space $K_{x,z}\in[10^{-1}, 10^3]$ for the same  dust abundance parameters as employed in section~\ref{sec:laminar} for the monodispse, MRN, an PB distributions.
These curves were obtained by  a local optimization in wavenumber space in steps across $\alpha$ to find the maximum growth rate within the space $K_{x,z}\in[10^{-1}, 10^3]$. Starting from the $\alpha=0$ limit with an initial guess derived from the wavenumber space grids in section~\ref{sec:laminar}, the growth rate was maximized at each step with a simplex algorithm search ({\tt scipy.optimize.minimize} with the Nelder-Mead algorithm) using the direct solver with 512 log-spaced points in $\taus$ space as the target function.
The direct solver \citepalias{paper1, paper2} was employed as it gives sufficient accuracy at the maximal growing point due to the second order convergence, but also provides a function which slopes towards that maximum from points in wavenumber space even where there is no physical growth (further discussed in Appendix~\ref{app:psidirectcomp}).
This conveniently prevents the simplex algorithm from failing when those points are tested in intermediate steps while searching for the maximum.

For the cases of most interest for planet formation, with $\mu > 1$, growth rates for either polydisperse distribution in Figure~\ref{fig:maximize_alpha}  are uniformly less than those for monodisperse dust.
In this linear analysis PSI does not save the streaming instability from the difficulties posed by the diffusive model of turbulent mixing.
However, it is also at least the case that the PB distribution maintains growth rates almost as large as those for monodisperse dust, so with dust processing able to create such distributions PSI should have similar planet forming potential to mSI even in the presence of turbulence.
Indeed, considering turbulence in a self-consistent manner should result in a stopping time dependent scale height for the dust, with greater settling for large grains than small.

In the $\mu=0.5$ cases shown in Figure~\ref{fig:maximize_alpha} the relation between the $\alpha$-dependency of the PSI and mSI is more complicated than in other scenarios.
We find that at $\tauspeak=10^{0}\Omega^{-1}$ the maximum growth rates scale similarly with $\alpha$ for the PB and monodisperse distributions.
Recall from Figure~\ref{fig:compmu0p5} how in this case at $\alpha=0$ the ridge of fast growth in wavenumber space was very similar for the PB and monodisperse distributions, and see here that it also damps with $\alpha$ in a very similar way.
At $\tauspeak=10^{-1}\Omega^{-1}$ the maximum growth rate for the monodisperse case is much faster in the $\alpha=0$ limit (Figure~\ref{fig:compmu0p5}) than with the MRN or PB distribution, and so the damping with increasing $\alpha$ is also quite different.
Only small islands of fast growth at low $K_z$ are present at $\alpha=0$ in the MRN and PB cases, and $\alpha$ needs to increase to significant level before affecting these.
In the monodisperse case, since the fastest growth at $\alpha=0$ is at high $K_x$ and $K_z$ the growth rates are damped much sooner as $\alpha$ is increased. Nevertheless, the maximum growth rates for the monodispserse case exceed the two PSI cases at all values of $\alpha$.
The surprising case in this $\mu=0.5$ set is the results for $\tauspeak=10^{-2}\Omega^{-1}$. Here, at $\alpha=0$ the maximum growth rate for the MRN and PB distributions was faster than for the monodisperse distribution (Figure~\ref{fig:compmu0p5}).
Since the instability at play is particularly well differentiated between the PSI cases (size resonance) and the monodisperse case (mSI) the effect of increasing $\alpha$ on the two is starkly different.
In the PSI cases, the increase in $\alpha$ gradually reduces the ridge of fast growing modes, from high-$K_z$ down.
For the mSI case, the high-$K_z$ section of the ridge of fastest growth dies away most quickly with increasing $\alpha$, while the low-$K_z$ section is more robust, leading to a decrease in growth in two parts.

We illustrate in detail of how the decrease in growth occurs with $\alpha$ in  Figure~\ref{fig:maximize_alpha}  for a typical case with $\mu=3$ and $\tauspeak = 10^{-1}\ \Omega^{-1}$ in
Figure~\ref{fig:alpha_growth_map}.
 Two basic trends are apparent in this series of wavenumber space growth maps.
First, increasing $\alpha$ results in damping of growth first at higher wavenumbers, 
and second, like in the inviscid case, the addition of the large particle enhancement in the PB distribution shifts not just the peak growth rates but the entire growth distribution in wavenumber space towards that of the monodisperse dust.
Second, the set of unstable modes present in the more slowly growing area abutting the $K_z$ axis is damped below the threshold first, before the ridge of fast growing modes. 
Recall in Figure~\ref{fig:mu3eignefuncs} that the difference between these two regions of growth in terms of the dust eigenfunctions was illustrated, and it was shown the the size resonance is at play in the PSI cases in the ridge of fast growth but not in the more slowly growing low-$K_x$ modes.
This difference appears to carry over to the viscous case too, with the different classes of modes having a different susceptibility to turbulent viscosity. 
This dependence on the type of mode extends to the relation between damping due to turbulent viscosity and damping due to dust diffusion. Damping due to dust diffusion contributes at least in similar magnitude to gas viscosity in all cases we investigated. The relationship between damping due to gas viscosity and dust diffusion is complex overall and depends at least on the dust distribution function and Schmidt number chosen and its dependence on particle size. This warrants further investigation.
Planetesimal formation is however unlikely to be strong affected by this property, as the low $K_x$ modes do not appear to show growth rates $\gtrsim10^{-2}\Omega$ in any case tested in this work.

Considering these results overall, it is apparent that, if this viscous diffusion model captures all of the effects of turbulence on the PSI or mSI, then the maximum $\alpha$ under which any form of SI could drive planetesimal formation in these conditions is a relatively low $\alpha \sim 10^{-5}$.
This suggests that the full nonlinear dynamics of the gas-dust system, beyond that captured here, may well be very important for enabling planet formation.
Stratified nonlinear three dimensional simulations of mSI self-consistently include particle-generated turbulence, which may behave differently than the simplified isotropic diffusion model of turbulence applied here and in similar calculations \citep{2019arXiv190605371U,2020ApJ...891..132C}.
Further development of numerical methods for 3D nonlinear fluid simulations with polydisperse dust will be required to tackle these questions.

\section{Discussion}
\label{sec:discussion}

\subsection{Scenarios for planet formation}
Armed with the results presented here, we are in the position for the first time to speculate on the impact of the real physical distribution of dust sizes on planetesimal formation, and hence planet formation.
It is observationally known that the dust size distribution in protoplanetary discs differs from that in the interstellar medium, so some degree of dust growth and a change to the shape of the dust distribution is known to occur.
To guide intuition, consider the simplest mapping of the canonical scenario for planetesimal formation by mSI to one based on PSI.
If fast linear growth of the instability (on the order of orbits) is a necessary condition for planetesimal formation, and the same underlying history of dust evolution is assumed, then beyond the single, peak size $\tauspeak=10^{-1}\ \Omega^{-1}$ we would expect a PB or PBT-like dust distribution of dust centered on this peak size.
In these conditions, the mSI grows quickly.
For particles with this stopping time, the mSI grows significantly faster for $\mu\gtrsim1$ \citep{2007ApJ...662..627J}.
However, the difference across this boundary is with the PSI is starker, so the requirement for $\mu\gtrsim1$ is stronger in the PSI case.

Producing this high local gas to dust ratio in a protoplanetary disc will require physically separating dust and gas by radial and vertical settling through the disc. 
Competing against this dust settling will be turbulent mixing and global gas flows.
Our results suggest that if the dust can concentrate to $\mu\gtrsim1$ then PSI can proceed on timescales only a few times longer than mSI.
However, to form a planetesimal, the PSI must grow to induce gravitational collapse of the concentrated dust cloud.
The conditions needed to produce this may not be the same, and may be more stringent, than those for fast linear growth.
It is possible that the PSI could be nonlinearly seeded by turbulent fluctuations in the dust concentration, violating the assumptions of the linear analysis used here.
Generally, we cannot say whether growing linear PSI modes will result in planetesimal formation via PSI. Nonlinear effects will likely dominate before PSI clumps can become massive enough to collapse under self gravity, so more barriers may need to be overcome in PSI planetesimal formation, and these barriers cannot be studied in the linear regime.
The analogous question for mSI may provide an imperfect guide.
It is difficult to interpret nonlinear mSI simulation results in this context, as although it has been shown that smaller ($\tauspeak=10^{-2}\ \Omega^{-1}$) dust is susceptible to mSI growth the resulting concentrations may have a weaker tendency to collapse into planetesimals \citep{2017ApJ...847L..12S,2019ApJ...883..192A,2020arXiv200801727C}.
This could be due to two separate reasons.
If this smaller dust is less prone to settling, then the $\mu$ which can be achieved is limited, and the mSI growth is only fast for this short stopping time dust when $\mu>1$. 
There does exist a $\mu<1$ regime where the PSI has a interesting property of supporting much faster growth then the mSI, but this appears to rely on an unevolved pure power-law dust distribution, and the fast growth is particularly susceptible to turbulent diffusion.
Achieving the right conditions for this kind of fast growth then appears to be very difficult.

Thus the most likely conditions for PSI driven planetesimal formation appear to be similar to those already found to produce self-gravitating fragments in nonlinear simulations of mSI.
Achieving the favourable conditions of $\mu\gtrsim1$ and $\tauspeak=10^{-1}\ \Omega^{-1}$ is not a trivial matter however.
For example, to produce  this regime at $45\ \mathrm{au}$ in the solar nebula \citet{2019NatAs...3..808N} invoked sub-cm pebbles under the conditions of a late-stage photoevaporating disc where the falling gas density lengthens the stopping time of a fixed size dust grain. 
Although this scenario is physically reasonable for a late-formed Kuiper belt object, the Solar System planets would need to form much earlier in a much denser gaseous disc -- particularly the gas giants, which need to acquire a gaseous envelope from the gaseous disc.

In an attempt to elucidate the conditions which appear to be favourable for PSI to drive planetesimal formation, we can appeal to simple disc models and limits for dust growth.
Assuming Epstein drag (typically applicable to disc conditions outside some tens of au) and a vertically isothermal disc structure, we can conveniently relate the dust particle radius to the stopping time as 
\begin{align}
\frac{\taus}{\Omega^{-1}} = \frac{a \rho_s}{\Sigma_{\rm g}} \frac{\pi}{2}\, ,
\end{align}
where $\Sigma_{\rm g}$ is the disc gas surface density \citep{2012A&A...539A.148B}.
Combining this with a minimum mass solar nebular disc model as is commonly used as a reference case for planet formation (also Appendix~\ref{app:bump}) yields
\begin{align}
r_{\rm fast} =11 \left(\frac{M_{\rm disc}}{M_{\rm MMSN}} \frac{\taus}{0.1\ \Omega^{-1}}\right)^{2/3}\left(\frac{a}{1\ {\rm cm}}\right)^{-2/3} \ {\rm au}\, ,
\end{align}
where $r_{\rm fast}$ is the radius where the criteria is achieved and $M_{\rm disc}/M_{\rm MMSN}$ is the ratio of the disc mass to a minimum-mass solar nebula (MMSN) \citep{2015A&A...579A..43C}.
Centimetre-scale dust in a MMSN thus has an $r_{\rm fast}=11\ {\rm au}$, and mm-scale dust an $r_{\rm fast}=51\ {\rm au}$.
Achieving $r_{\rm fast}=5\ {\rm au}$ requires dust with radius $\sim3\ {\rm cm}$, and  $r_{\rm fast}=1\ {\rm au}$ a dust radius of  radius $\sim40\ {\rm cm}$, although in a MMSN these latter two cases are entering into the Stokes drag regime.
If the gas disc is denser than an MMSN model, these radial locations are pushed outwards, and the required dust sizes down.

To estimate the sizes of the largest dust aggregates present in the disc, we can appeal to the model of \citet{2012A&A...539A.148B}, and in a similar MMSN-type disc (Appendix~\ref{app:bump}) obtain for the stopping time of the fragmentation-limited dust aggregates $\tau_{\rm s, frag}$:
\begin{align}
\frac{\tau_{\rm s, frag}}{\Omega^{-1}} = 0.023\left(\frac{u_{\rm f}}{1\ {\rm m/s}}\right)^2 \left(\frac{r}{1\ {\rm au}}\right)^{1/2}\left(\frac{\alpha}{10^{-5}}\right)^{-1}
\end{align}
where $u_{\rm f}$ is the dust aggregate fragmentation velocity, the turbulence parameter $\alpha$ sets the scale of the velocities relative to the sound speed of the vortices driving the collisions.
This fragmentation velocity is the same as appears also in the physics leading to the PB and PBT dust distributions as described in Appendix~\ref{app:bump}. 
The appearance of the turbulence parameter $\alpha$, previously also present in the model for the diffusive effects of turbulence (Section~\ref{sec:turbmodel}), is due to the role of turbulent eddies in driving the equal-size dust collisions in \citet{2012A&A...539A.148B}. 
A major weakness of this common manner of parameterizing the physics is the likely possibility that the complexity of turbulent phenomena is beyond a single scalar parameter, and such a tight linkage between the diffusive effect and dust collision driving may not be an appropriate model for the underlying physics.
Nevertheless, in this model and at the turbulence level of $\alpha=10^{-5}$ at $1\ {\rm au}$,   $\tau_{\rm s, frag}=0.1 \ \Omega^{-1}$ is not achieved with a fragmentation velocity 
$u_{\rm f}=1{\rm\ m/s }$, but if it is $10\ {\rm m/s}$ the required stopping time is easily achieved.
However, if we take the diffusive model of turbulence at face value, considering the results of Section~\ref{sec:turbmodel} the value of $\alpha=10^{-5}$ is a significant impediment to the growth of PSI (or mSI).
Instead considering a very low value of $\alpha=10^{-6}$ then $u_{\rm f}=1{\rm\ m/s }$ is sufficient for producing $\tau_{\rm s, frag}=0.1 \ \Omega^{-1}$  at $1\ {\rm au}$ in this model.
If  planetesimals can be formed from dust with lower peak stopping times (lower Stokes number) then the role of the dust enhancement at the peak particle size may be lessened. With $\tauspeak=10^{-2}\ \Omega^{-1}$ we found growth at least on the $10^{-2}\ \Omega$ level, in all the laminar cases in this work with the MRN distribution, unlike at longer peak stopping times. This may open a different window of opportunity for planetesimal formation.
This simplified discussion serves to illustrate the difficulty of obtaining the fuel for PSI driven planetesimal formation in the disc, but being simplified it has also omitted many detailed considerations that could still be significant.

Other hydrodynamic dust enrichment and size sorting pathways are thought to occur in protoplanetary discs. Notably, vortices in protoplanetary discs have pressure peaks at their cores which collect dust \citep{1995A&A...295L...1B,2004_johansen}. Larger dust particles collect in the vortex core faster than smaller particles causing a degree of size sorting as well as dust concentration. The lifetimes of these dust loaded vortices has been called into question due to instabilities shown in \citet{Surville_2019} and \citet{DVorticesII_inprep}, though both works agree that, at least in 2D studies, the vortices survive long enough to considerably increase dust concentration in their cores before succumbing to instabilities. Furthermore \citet{DVorticesII_inprep} finds small vortices to be stable for even longer periods, of the order of hundreds of orbits, giving them plenty of time to cause a considerable dust enrichment at their cores.

Turbulent flow in disc gas can be expected to size-sort and concentrate dust, producing both a localized concentration of dust and a lognormal distribution (in stopping time) with the peak located where the particle stopping time equals the Kolmogorov scale eddy turnover time
\citep{2001PhFl...13.2938H}
Depending on the turbulence parameters, this size scale can be appropriate for concentrating chondrule-sized particles
\citep{2001ApJ...546..496C}, but these are typically a few orders of magnitude below the favourable $\taus=0.1\ \Omega^{-1}$ identified for PSI.

As physical evidence that a PB-like dust distribution may occur in some circumstances in a planet-forming disc, one can appeal to the size-sorting of chondrules and other components of chondritic meteorites
\citep{2010M&PS...45.1124T}
Models for the distributions vary, but 
chondrule size distributions have relatively narrow peaks with widths often similar or narrower than the PB distribution considered here \citep{2015ChEG...75..419F,2018M&PS...53.1489M}.
Chondritic meteorites, between the chondrules and other inclusions, consist of a fine-grained matrix material.
The grains in this material often have a power-law size distribution, with for example the Allende CV chondrite matrix grains having a top-heavy power law distribution with $\beta\sim -2$
\citep{1977E&PSL..35...25A,1989E&PSL..92..265T}.
That some chondrites have very low fractions of matrix grains and a minimum chondrule size does however suggest an efficient size-sorting process, although one which may be a result of accretion through an asteroid atmosphere and not related to dust concentration in the disc itself
\citep{2015SciA....1E0109J,2019ApJ...887..230M}. 

The reproduction of the inclination distribution of binary orbits in the Kuiper belt from mSI simulations is perhaps the strongest argument a the present time for the involvement of SI in the formation of a solar system \citep{2019NatAs...3..808N}.
Simulations in that work tested formation from particle with stopping times of either $0.2\ \Omega^{-1}$ or $2\ \Omega^{-1}$.
Our results suggest that the only modification to their scenario required to produce similar results with the PSI is the dust size evolution or sorting to produce a PB distribution-like (or more severe) abundance of the largest grains in the dust distribution over an MRN power-law distribution. 
Testing this will require a new generation of numerical methods to perform nonlinear PSI simulations.

If however, dust evolution models are unable to produce sufficiently large dust with a sufficiently narrow size distribution at sufficient concentration in the disc for fast PSI growth to occur, alternate models of planetesimal formation, such as direct turbulent concentration \citep{2020ApJ...892..120H}, may provide a path to planet formation.

\subsection{Limitations and future directions}
Our study has been limited in various ways.
We have only considered Epstein drag in this work and the preceding two \citepalias{paper1,paper2}.
Particles with $\tau_{\rm s, frag}\sim 0.1 \ \Omega^{-1}$  at $1\ {\rm au}$ in an MMSN ought to lie in the Stokes drag regime.
However, \citet{2020ApJ...891..132C} did not find significant effects in the linear mSI from the inclusion of Stokes drag, and this property may carry over to the PSI.

The aspect of the PB and PBT dust distribution which appears to allow enhanced PSI growth over a pure power-law MRN distribution is the `cratering bump' attributed to the effect on coagulation-fragmentation from the outcome of differing-sized dust collisions included in the \citet{2011A&A...525A..11B} model.
In the denser parts of the disc, aeolian erosion may act to limit the size of weakly bonded aggregates \citep{2020MNRAS.496.4827R,2020ApJ...898L..13G}.
A coagulation-fragmentation model including this effect should be used in the future to determine where this alters the appearance of the cratering bump, and the methods used here employed to test the effect on PSI growth.

Radially varying  gas pressure bumps in the disc may slow the radial drift of dust and cause a size-dependent local enhancement in the dust density.
As \citet{2020arXiv200801727C} have proposed that these dust density enhancements due to mild pressure bumps in the disc can promote fragmentation through mSI, the effect of pressure bump and the changes to the dust distribution due to size-dependent traffic-jam effects may have an effect on the local dust size distribution along with the dust to gas ratio, and hence an effect on determining the growth of PSI.
Locally in pressure bump, globally across the radii of a given protoplanetary disc, and in other contexts where PSI may occur, such as lunar formation \citep{2020LPI....51.2976A}, the radial pressure gradient $\eta$ may vary, and in particular be smaller than the value used in this survey.
The effect of this warrants further investigation in both linear and nonlinear studies of the PSI.

We have only considered vertically unstratified models. Linear mSI calculations for the stratified case have only appeared very recently \citep{2020arXiv201112300L}. It will be very interesting to extend these calculations to the polydisperse case. 

The results on turbulent PSI in this work are also limited in several ways. First of all, we consider only the gas turbulence to dust stirring relation described in equation \ref{eqn:dust-turbulence-coupling}, which allows easy comparison with previous work done on the mSI. The character of the turbulent PSI is however strongly dependant on the relationship between dust stirring and turbulence, as dust diffusion plays an important part in damping growth in turbulent PSI. In practice what this means is that the way different dust distributions are affected by turbulence ultimately depends very strongly on the relationship between gas turbulence and dust diffusion at different dust sizes. Secondly, we neglect the effects of dust on turbulence. This may have effects on the relationship between dust and turbulence when considering more complete models of dusty turbulence. Overall, further dedicated study of turbulent PSI may yield interesting results if a more self-consistent approach to modelling  turbulence is adopted.

The timing and location of planet formation driven by PSI must be dependent on the dust evolution and concentration that leads to fast PSI growth, and so a model  of these timescales and locations should encompass a disc-scale dust evolution model.
In addition, size-differential dust concentration by PSI should also be investigated
as it could directly alter the input to dust coagulation-fragmentation processes, 
or lead directly to new and different initial conditions for further stages of streaming instability.

Finally, this work and the previous ones in the series have only considered linear stability.
In our linear calculations, we found that the dust which concentrates most strongly is not the largest (i.e.~$\taus$ closest to unity) dust as one might naively expect from intuition based on the mSI, but instead the dust near the size-resonance in modes where it dominates.
If this behaviour persists into the nonlinear stage of the instability the expectation  for the constituent particle sizes of bodies formed from PSI may not be a simple reflection of the particles which had $\taus$ closest to unity in the formation environment. 
Of course, planetesimal formation results from the gravitationally induced collapse of a dust overdensity.
To fully test the ability of the PSI to form planetesimals, nonlinear simulations including self-gravity will be required.
The results of these papers, and the numerical tools released will serve as useful benchmarks for the ability of those simulation codes to capture the PSI.

\section{Conclusions}
\label{sec:conclusions}
We have performed a first survey of the conditions and linear growth rates of the PSI likely to be encountered in the planet formation process.
We confirm that wide MRN top-heavy power-law dust distributions result in damped growth of PSI in  the most regimes unstable for the mSI, with an exception occurring for short stopping times and dust to gas ratios of order unity.
The faster growth in these cases is confined to relatively short wavelengths, and so is vulnerable to disruption by other effects.
However, an enhancement of large particles, like that produced by reasonable dust evolution models, allows linear PSI growth nearly as fast the mSI, as long as the dust to gas ratio $\mu\gtrsim 1$.

We have shown that the theory and tools developed in this series of papers is capable of determining if a streaming instability will grow quickly for a given set of dust parameters, and for the first time taking accurately into account an arbitrary dust size distribution.
This provides the basis for solving one of the major unknowns with the proposed role of streaming instabilities in planet formation - the nature of the dust evolution required to allow the instability to proceed.
As streaming instabilities require at least some dust to grow to sizes with stopping times on the order of an orbit for fast growth, the issue of to what extent the largest dust needs to grow, and how dominated by large dust (`top-heavy') the distribution needs to be must be addressed.
Our results enable the outcome of dust evolution processes to be tested to determine if they give rise to fast PSI.

From the examples studied in this work, we are able to conclude that if fast linear growth of the PSI is a prerequisite for planet formation, then one favourable scenario relies both on production of dust of sufficiently large stopping times ($\tauspeak=10^{-1}\ \Omega^{-1}$), concentration of dust to dust to gas ratio of at least order unity ($\mu>1$) and a dust evolution to a dust size distribution enhanced above canonical top-heavy power-law distributions inferred from pure fragmentation cascade dust evolution the interstellar medium.

We find that a diffusive model of turbulence has similar impact on the linear PSI as it does to the linear mSI, and so requires similar testing in nonlinear and self-consistent simulations.
These linear results need to be further tested with nonlinear simulations, allowing the examination of the saturation of PSI and the collapse of fragments from the enhanced density dust clouds formed.

Finally, though this survey has been inherently limited, the underlying tools \citep{psitools} have been made available to the community so that similar analysis can be performed for specific predicted dust distributions arising from dust evolution models.

\section*{Acknowledgements}

We thank the anonymous referee for insightful and constructive comments.
We acknowledge useful discussions with Richard Nelson, Chao-Chin Yang, and Mordecai-Mark Mac~Low.
This research was supported by an STFC Consolidated grants awarded to the QMUL Astronomy Unit 2017--2020 ST/P000592/1 and 2020--2023 ST/T000341/1.
We acknowledge that the results of this research have been achieved using the DECI resource Beskow based in Sweden at PDC with support from the PRACE aisbl.
This research utilised Queen Mary's Apocrita HPC facility, supported by QMUL Research-IT \citep{apocrita}.
SJP is supported by a Royal Society URF.


\section*{Data availability}

The software used to perform calculations in this work is publicly archived on Zenodo.org \citep{psitools}.



\bibliographystyle{mnras}
\bibliography{psi2} 




\appendix

\section{Typical Cratering Bump Parameters}
\label{app:bump}

\begin{figure}
\centering
	\includegraphics[width=0.8\columnwidth]{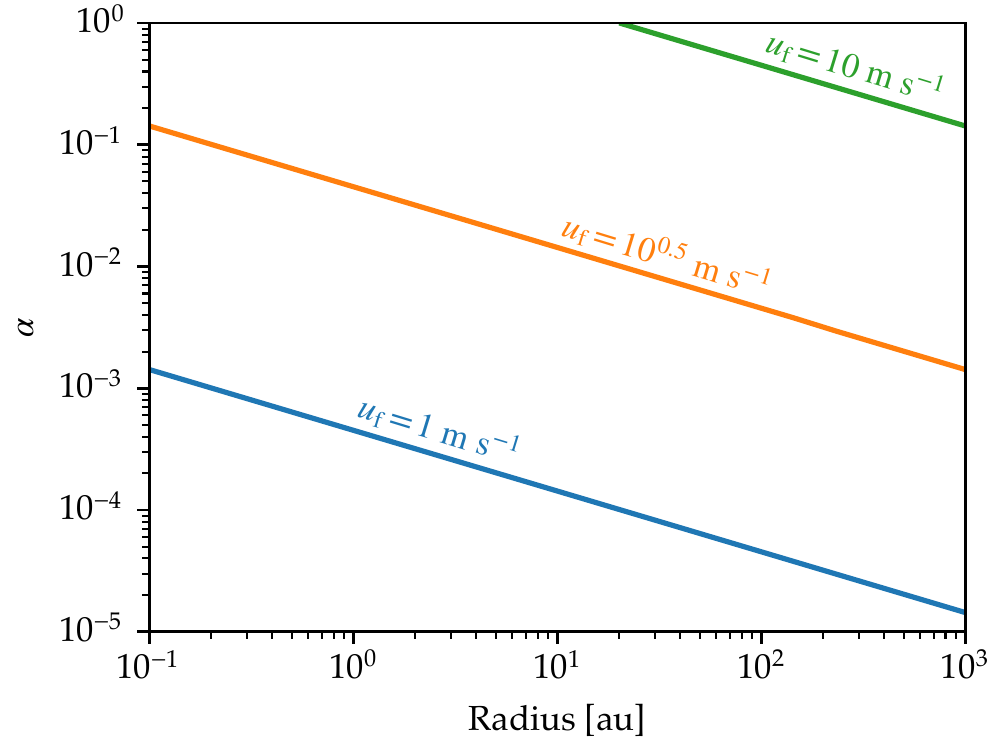}
    \caption{Upper limits on the range where the \citet{2011A&A...525A..11B} recipe parameters are the values used in the PB distribution for representative values of the aggregate fragmentation velocity $u_{\rm f}$.}
    \label{fig:pblimits}
\end{figure}

\begin{figure}
\centering
	\includegraphics[width=1.0\columnwidth, trim=2cm 0 1cm 0, clip]{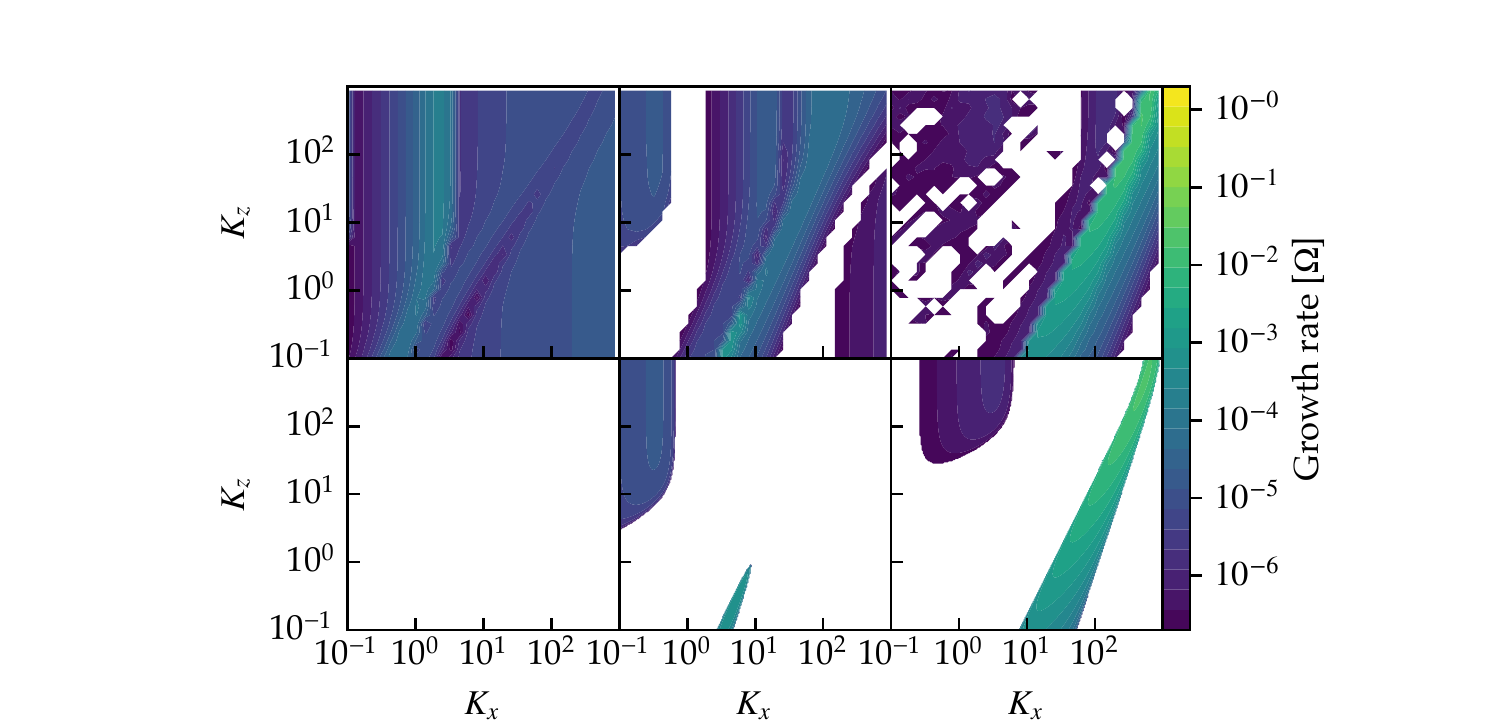}
	\includegraphics[width=1.0\columnwidth, trim=2cm 0 1cm 0, clip]{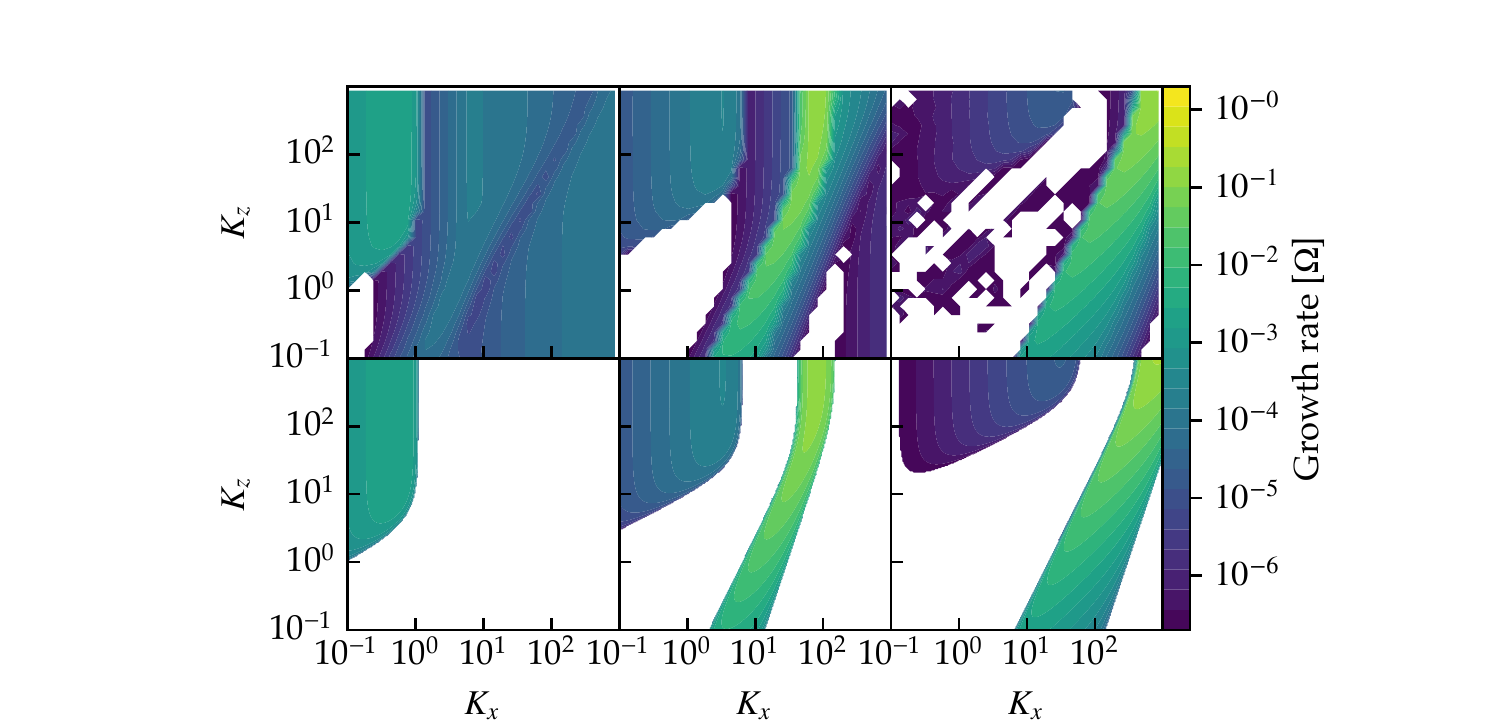}
	\includegraphics[width=1.0\columnwidth, trim=2cm 0 1cm 0, clip]{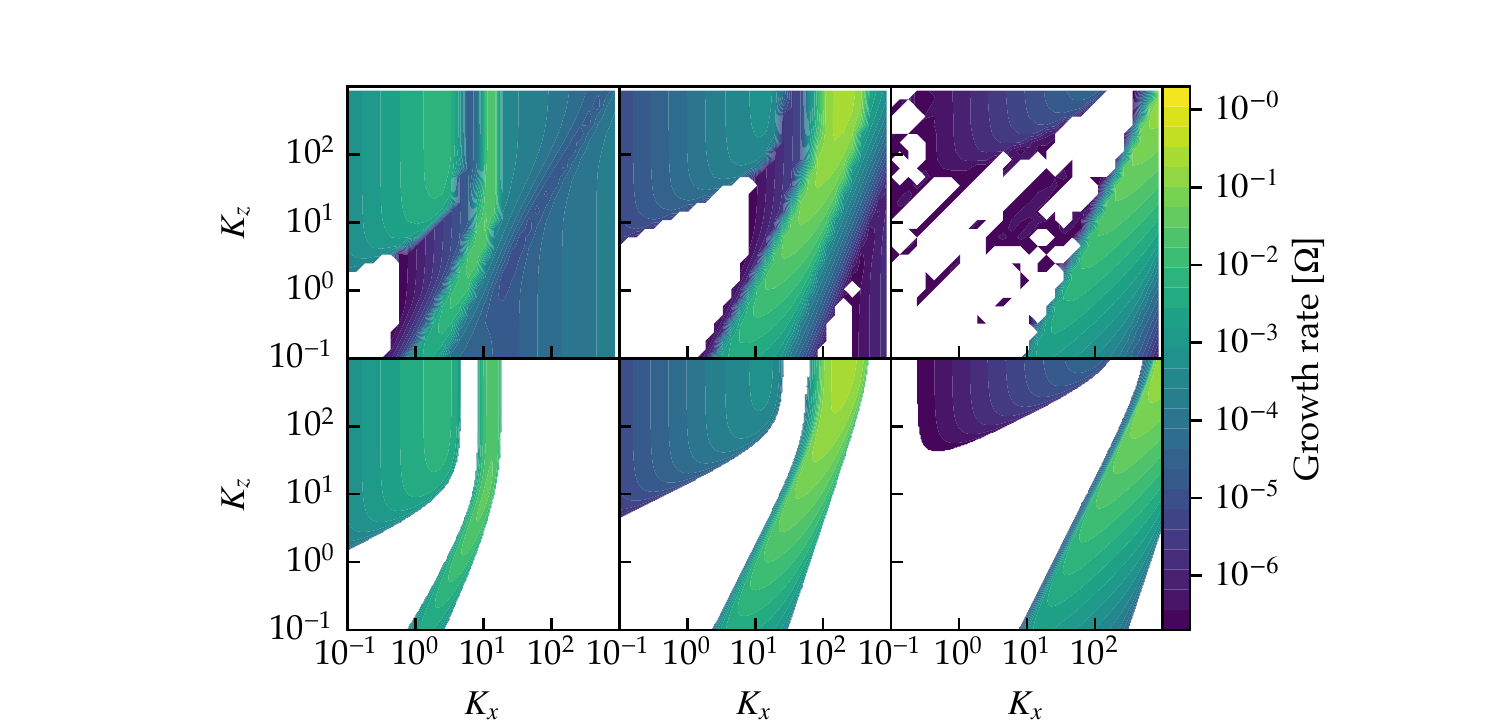}
    \caption{Comparison of direct solver and root finder results for MRN dust distributions with 
    {\sl Top Panel}: $\mu=0.5$
    {\sl Middle Panel}: $\mu=3$
    {\sl Lower Panel}: $\mu=10$
    {\sl Top Rows}: direct solver
    {\sl Bottom Rows}: root-finder grid-refinement map
    {\sl Left Column}: $\tauspeak=10^{0}\Omega^{-1}$
    {\sl Middle Column}: $\tauspeak=10^{-1}\Omega^{-1}$
    {\sl Right Column}: $\tauspeak=10^{-2}\Omega^{-1}$
    }
    \label{fig:dpsimrn}
\end{figure}

\begin{figure}
\centering
	\includegraphics[width=1.0\columnwidth, trim=2cm 0 1cm 0, clip]{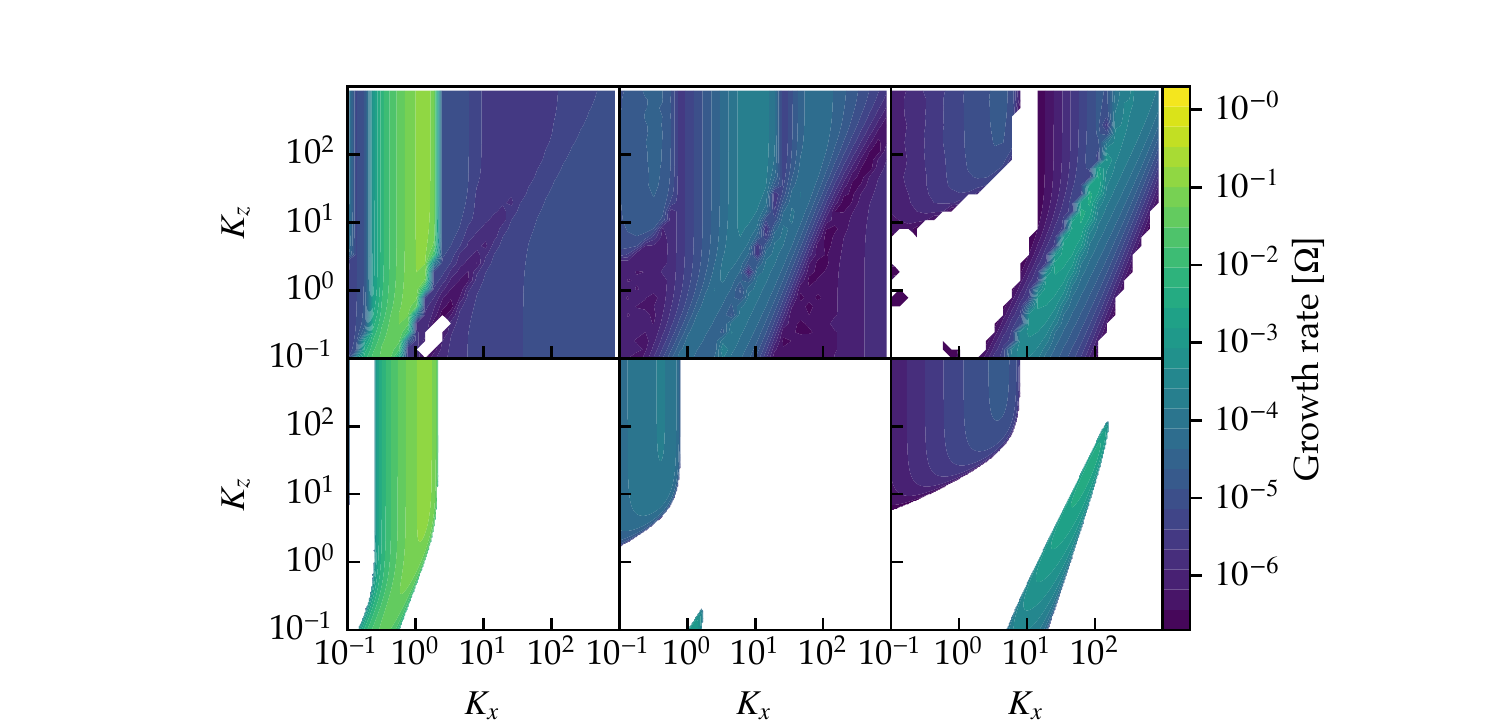}
	\includegraphics[width=1.0\columnwidth, trim=2cm 0 1cm 0, clip]{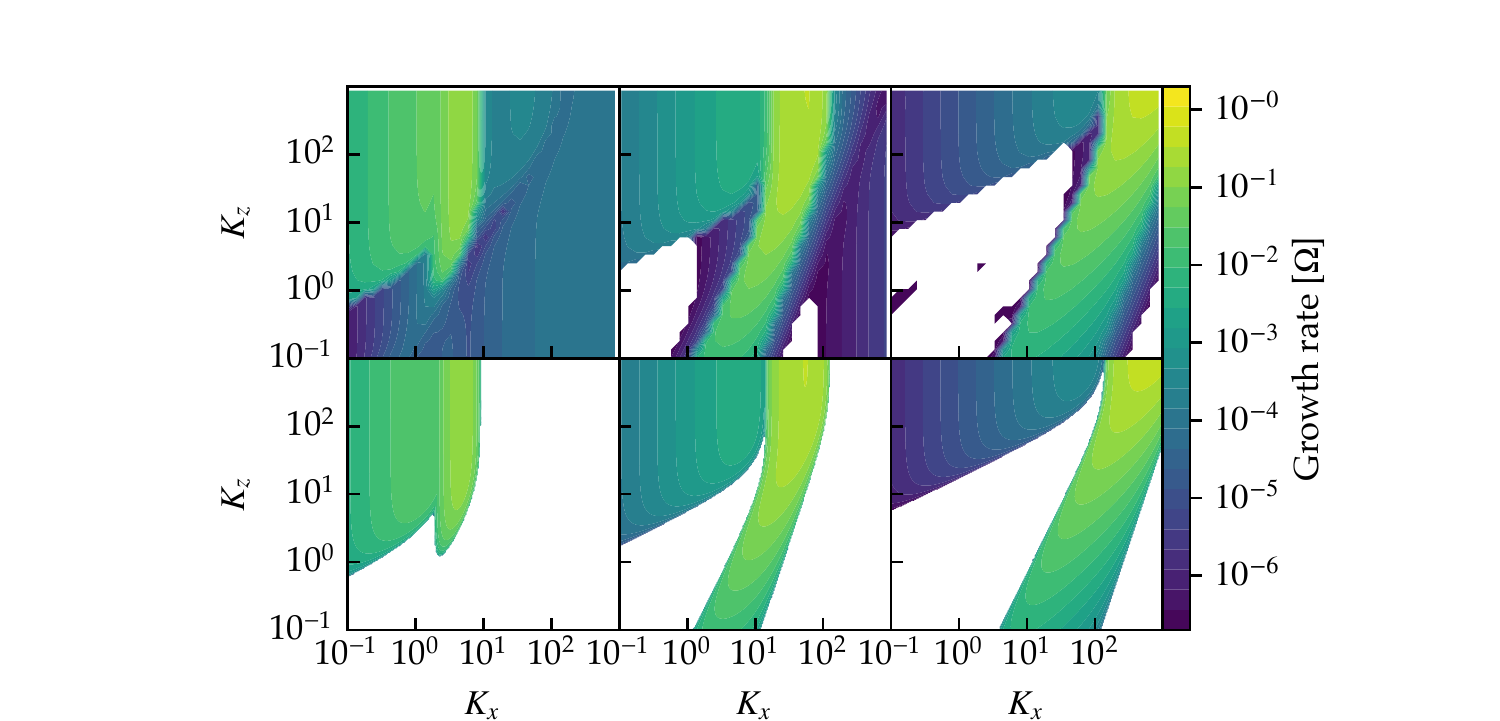}
	\includegraphics[width=1.0\columnwidth, trim=2cm 0 1cm 0, clip]{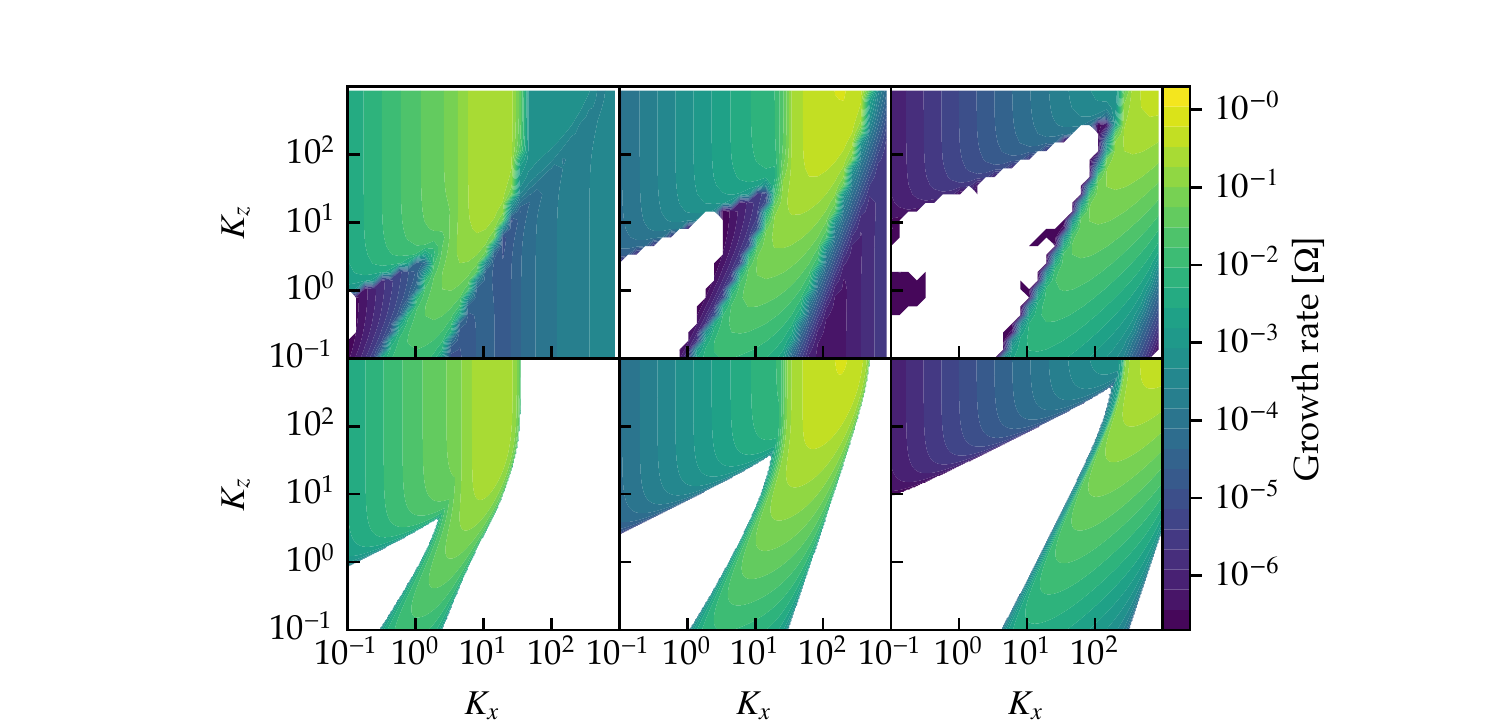}
    \caption{Comparison of direct solver and root finder results for PB dust distributions with 
    {\sl Top Panel}: $\mu=0.5$
    {\sl Middle Panel}: $\mu=3$
    {\sl Lower Panel}: $\mu=10$
    {\sl Top Rows}: direct solver
    {\sl Bottom Rows}: root-finder grid-refinement map
    {\sl Left Columns}: $\tauspeak=10^{0}\Omega^{-1}$
    {\sl Middle Columns}: $\tauspeak=10^{-1}\Omega^{-1}$
    {\sl Right Columns}: $\tauspeak=10^{-2}\Omega^{-1}$
    }
    \label{fig:dpsiPB}
\end{figure}

\begin{figure*}
	\includegraphics[width=0.8\textwidth]{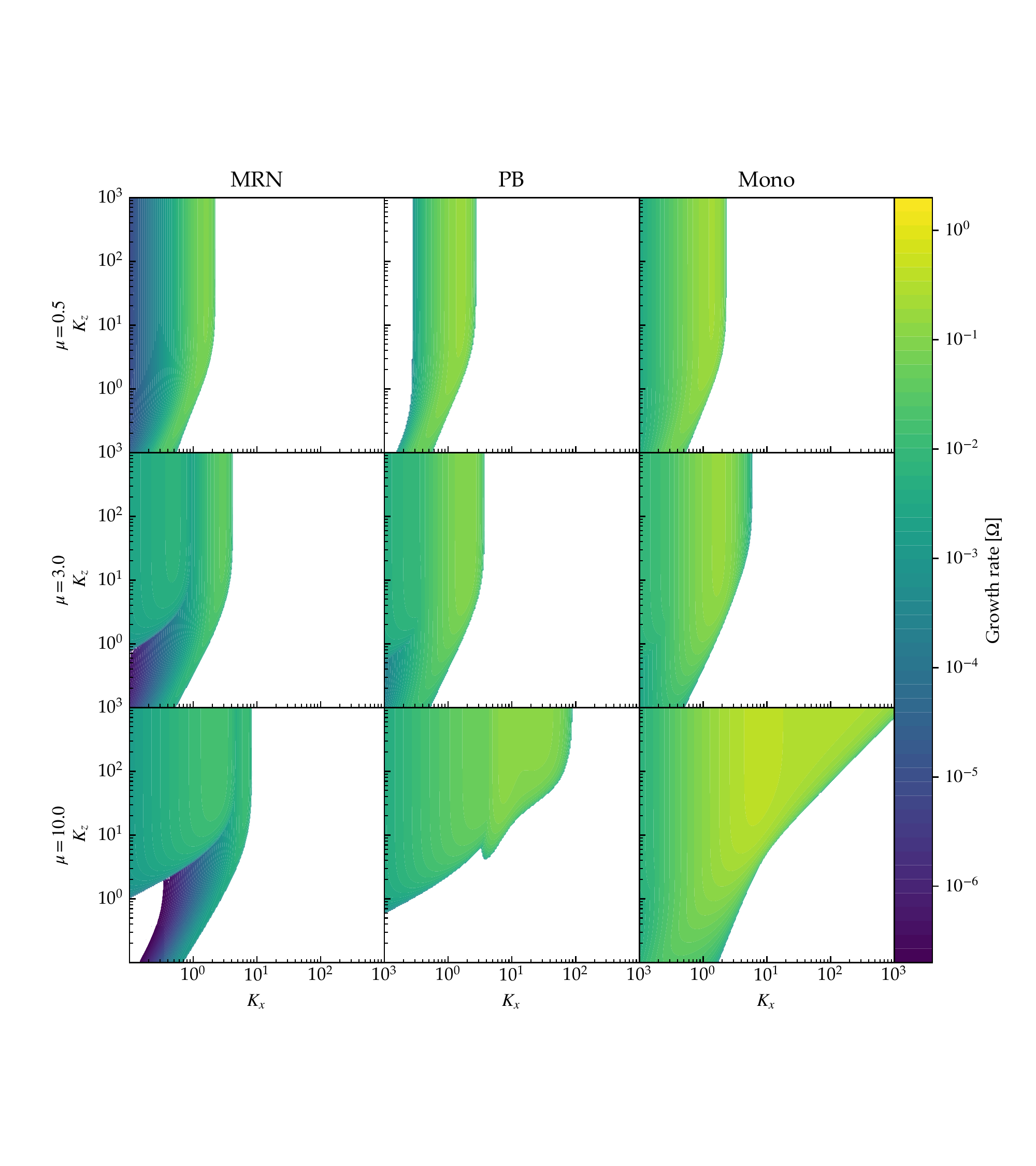}
    \caption{PSI growth rate maps for MRN, PB, and monodisperse distributions with $\tauspeak=3$, and $\mu\in \{0.5, 3, 10\}$.}
    \label{fig:longstop}
\end{figure*}

This appendix gives detail on  the process used to choose the parameters used for the cratering bump in the PB and PBT distributions.
Although the \citet{2011A&A...525A..11B}  allows for a range of values, calculated from basic parameters, this work only considers the particular values given in Equations~(\ref{eq:aL}) and~(\ref{eq:aR}).
Evaluating the expressions for the parameters $a_{\rm L}$, $a_{\rm P}$, and $a_{\rm R}$ of \citet{2011A&A...525A..11B} in a disc with surface density $\Sigma$ and sound speed $c_{\rm s}$ specified as: 
\begin{align}
    \Sigma = 1700\ {\rm g\ cm^{-2}}\ \left(\frac{r}{1\ {\rm au}}\right)^{-3/2}\\
    c_{\rm s} = 1.2\ {\rm km\ s^{-1}}\ \left(\frac{r}{1\ {\rm au}}\right)^{-1/4}
\end{align}
with monomers of size  $a_0=0.025\ {\rm \mu m}$ and solid density $\rho_s=3\ {\rm g\ cm^{-1}}$
yields constant ratios of the cratering bump parameters, namely $a_{\rm L}/a_{\rm P}=2/3$  and $a_{\rm R}/a_{\rm P}=1.5625$ (or $25/16$), for most regions of the disc, below an upper limit on on the turbulence parameter $\alpha$.
These curves, as a function of disc radius $r$ and turbulence parameter $\alpha$, are shown in Figure~\ref{fig:pblimits} for three representative values of the dust aggregate fragmentation velocity $u_{\rm f}$.
Experimental data suggests the fragmentation velocity for silica aggregates can be the order of $1\ {\rm m\ s^{-1}}$, and possibly $10\ {\rm m\ s^{-1}}$ for water ice aggregates \citep{2000ApJ...533..454P,2015ApJ...798...34G,2020MNRAS.497.2517B}.
In practice we do not assign a deeper physical significance to the numerical value these parameters take (and truncate $a_{\rm R}/a_{\rm P}$ to $1.56$ for convenience), as the \citet{2011A&A...525A..11B} model is a semianalytic fit to a much more sophisticated and self-consistent set of calculations. 
The limits Figure~\ref{fig:pblimits} in do however give a guide to where the expected dust distribution properties ought to significantly deviate from these values in that model. 
For study of more general and sophisticated dust distributions in linear PSI, the publicly available {\tt psitools} package \citep{psitools} can be used.


\section{Direct and root finding eigenvalue map comparison}
\label{app:psidirectcomp}

In this appendix we compare the results of high resolution direct solver results to those from the root finder for some of the main cases studied in this paper.
The root-finding algorithm for finding PSI eigenvalues benefits from 
excellent speed and precision, but is inherently stochastic. 
As a control on the possible omissions of growing modes made by this, we present here equivalent calculations obtained with the direct solver
with 4096 points in $\taus$ space and Chebyshev roots gridding, on a more sparsely sampled wavenumber space grid.
Figure~\ref{fig:dpsimrn} corresponds to the cases in the top rows of Figures~\ref{fig:compmu0p5}--\ref{fig:compmu10}, and
Figure~\ref{fig:dpsiPB} corresponds to the cases in the middle rows of Figures~\ref{fig:compmu0p5}--\ref{fig:compmu10}.
In general the fastest growing modes are captured very well with both solvers, but the regions of much slower growth are poorly handled by the direct solver, as explained in \citetalias{paper1} and \citetalias{paper2}. 
Importantly, these maps verify that no island of fastest growing modes has been missed by the root finder in these cases, building confidence in the correctness of the results.
As explained in \citetalias{paper2} both larger errors and first order convergence can be expected with the direct solver where no growing mode exists, and so these regions often have a value at the 4096 point resolution used where which is above the $2\times10^{-7}\Omega$ threshold.

The direct solver results also illustrate why the direct solver is more useful than the root finder for use with a simplex optimization of the growth rate over wavenumber. For example, in the upper panel of Figure~\ref{fig:dpsiPB} for $\mu=0.5$ $\tauspeak=10^{-2}\Omega^{-1}$ the direct solver result provides a smooth slope towards the fasest growth rate, even outside of the region where the root finder easily shows there is no growing mode. This allows the slope-following habit of the simplex algorithm to correctly find the fastest growth rate even when starting  at a wavenumber where the is no appreciable growth.

\section{Long stopping times}
\label{app:longstop}

The self-consistency of the governing equations has consistency problems for $\tauspeak \gtrsim 1\ \Omega^{-1}$ as discussed in  \citetalias{paper1} and Appendix~A of \citetalias{paper2} since the fluid approximation breaks down \citep{2011MNRAS.415.3591J} and particles fail to settle smoothly to the midplane \citep{1995Icar..114..237D}.
Notwithstanding this, the mathematical stability eigenproblem is well posed and can still be solved.
For sake of understanding in relation to \citet{2020arXiv200801119Z}, this appendix includes calculations in this long stopping time regime.
Figure~\ref{fig:longstop} shows wavenumber space growth maps for $\tauspeak=3\ \Omega^{-1}$ for $\mu \in \{0.5, 3, 10\}$.

These maps have a base grid of $16\times 16$ with 4 levels of refinement, except for the MRN $\tauspeak=3\ \Omega^{-1}$ case which uses a finer $32\times 16$ grid, with a additional higher resolution patch in the lower right corner $K_x,\ K_z \in [0.1, 0.34145]$ with a base grid $128\times16$ and three levels of refinement.
This was found to be necessary to reliably locate the roots with very small growth rates.

In the $\mu=0.5$ case in Figure~\ref{fig:longstop} a notably different trend is present form that in the $\tauspeak=1\ \Omega^{-1}$ data from Figure~\ref{fig:compmu0p5}.
In the $\tauspeak=1\ \Omega^{-1}$ case, the MRN distribution map showed no appreciably growing modes at any wavenumber, and then the same growth pattern emerging  (a $K_x \sim K_z$ ridge transitioning to a $K_z \sim \text{const}$ ridge) in the PB, PBT, and monodisperse cases.
However, here with $\tauspeak=3\ \Omega^{-1}$ the MRN case displays a ridge of fast growth at roughly the same location as that in the the PB and monodisperse cases.
This  $\tauspeak=3\ \Omega^{-1}$ is wildly outside of the short stopping time regime analyzed with the terminal velocity approximation in \citepalias{paper1}, and the pattern shown here suggests that in this long stopping time regime the nature of the instability at low-$\mu$ changes.
That growth occurs here with a MRN distribution at $\mu\lesssim 1$ but with $\tauspeak>3\ \Omega^{-1}$ is in agreement with the conclusions of \citet{2020arXiv200801119Z}.


\bsp	
\label{lastpage}
\end{document}